\documentclass[twocolumn]{aastex631}

\begin{document}


\title{Multiple Components of the Outflow in the Protostellar System HH 212:\\
Outer Outflow Shell, Rotating Wind, Shocked Wind, and Jet}
\shortauthors{L\'opez-V\'azquez et al.}

\author[0000-0002-5845-8722]{J. A. L\'opez-V\'azquez}
\thanks{Corresponding author: J. A. L\'opez-V\'azquez}
\email{jlopezv@asiaa.sinica.edu.tw}
\affiliation{Academia Sinica Institute of Astronomy and Astrophysics, No. 1, Sec. 4, Roosevelt Road, Taipei 10617, Taiwan}

\author[0000-0002-3024-5864]{Chin-Fei Lee}
\affiliation{Academia Sinica Institute of Astronomy and Astrophysics, No. 1, Sec. 4, Roosevelt Road, Taipei 10617, Taiwan}

\author[0000-0001-8385-9838]{Hsien Shang}
\affiliation{Academia Sinica Institute of Astronomy and Astrophysics, No. 1, Sec. 4, Roosevelt Road, Taipei 10617, Taiwan}

\author[0000-0002-1593-3693]{Sylvie Cabrit}
\affiliation{Observatoire de Paris, PSL University, Sorbonne University, CNRS, LERMA, 75014 Paris, France}
\affiliation{Univ. Grenoble-Alpes, CNRS, IPAG, 38000 Grenoble, France}

\author[0000-0001-5557-5387]{Ruben Krasnopolsky}
\affiliation{Academia Sinica Institute of Astronomy and Astrophysics, No. 1, Sec. 4, Roosevelt Road, Taipei 10617, Taiwan}

\author[0000-0003-1514-3074]{Claudio Codella}
\affiliation{INAF, Osservatorio Astrofisico di Arcetri, Largo E. Fermi 5, I-50125 Firenze, Italy}

\author[0000-0002-1624-6545]{Chun-Fan Liu}
\affiliation{Academia Sinica Institute of Astronomy and Astrophysics, No. 1, Sec. 4, Roosevelt Road, Taipei 10617, Taiwan}

\author[0000-0003-2733-5372]{Linda Podio}
\affiliation{INAF, Osservatorio Astrofisico di Arcetri, Largo E. Fermi 5, I-50125 Firenze, Italy}

\author[0000-0002-2338-4583]{Somnath Dutta}
\affiliation{Academia Sinica Institute of Astronomy and Astrophysics, No. 1, Sec. 4, Roosevelt Road, Taipei 10617, Taiwan}

\author[0000-0003-0376-6127]{A. Murphy}
\affiliation{Academia Sinica Institute of Astronomy and Astrophysics, No. 1, Sec. 4, Roosevelt Road, Taipei 10617, Taiwan}

\author[0000-0002-1143-6710]{Jennifer Wiseman}
\affiliation{Exoplanets and Stellar Astrophysics Laboratory, NASA Goddard Space Flight Center, Greenbelt, MD 20771, USA}



\begin{abstract}
We present the Atacama Large Millimeter/submillimeter Array Band 7 observations of the CO ($J=3-2$) line emission of the protostellar system HH 212 at $\sim$24 au spatial resolution and compare them to those of the SiO ($J=8-7$) and SO ($J=8-7$) line emission reported in the literature. 
We find that the CO line traces four distinct regions: (1) an outer outflow shell, (2) a rotating wind region between the SiO and CO shells, (3) the shocked and wide-angle inner X-wind inside a SiO shell, and (4) the jet. The origin of the CO outer outflow shell could be associated with the entrained material of the envelope, or an extended disk wind. The rotating wind, which is shocked, is launched from a radius of $9-15\,\mathrm{au}$, slightly exterior to that of the previously detected SO shell, which marks the boundary where the wide-angle X-wind is interacting with and shocking the disk wind. Additionally, the SO is found to be mixed with the CO emission within the thick and extended rotating wind region. The large scale CO shocked wind coexists with the SO emission near the upper portion of the inner shocked region converged on top of the inner SiO knots. The CO jet is traced by a chain of knots with roughly equal interval, exhibiting quasi-periodicity, as reported in other jets in the literature.
\end{abstract}



\section{Introduction}
\label{sec:introduction}

The earliest phases of the star formation process (Class 0/I, with ages of $\leq10^4$ yr and $\sim10^5$ yr, respectively, \citealt{Andre2000}) are dominated by high accretion of material from the envelope to the protostar-disk system (\citealt{Hartmann1996}). As a result of the conservation of angular momentum due to the accretion, the protostar-disk system ejects material from the inner region of the disk in the form of jets and bipolar outflows. Although the jets and outflows are predominantly observed in young protostellar systems (Class 0/I), e.g., the Herschel Orion Protostar Survey (HOPS; \citealt{Furlan2016}; \citealt{Hsieh2023}), ALMA Survey of Orion Planck Galactic Cold Clumps (ALMASOP; \citealt{Dutta2020}; \citealt{Dutta2024}), the Perseus ALMA Chemistry Survey (PEACHES; \citealt{Yang2021}), and  CALYPSO (\citealt{Podio2021}), several jets and outflows associated with more evolved systems (Class II, with an age of $\sim10^6$ yr, \citealt{Andre2000}) have also been observed, such as IRAS 16316-1540, T Tau, and GK /GI Tau (\citealt{Arce2006}), DO Tauri (\citealt{FernandezLopez2020}), and HH 30 (\citealt{Pety2006}).

Both jets and outflows may be responsible for removing excess angular momentum from the protostar-disk system (\citealt{Ray2007}). This angular momentum is carried away by the vertical magnetic field lines anchored to the disk by the magneto-centrifugal mechanism (\citealt{Blandford1982}). The magnetic field lines can be anchored in two regions: close to the central star, where the stellar magnetosphere truncates the disk, known as X-winds (\citealt{Shu1994}; \citealt{Shang1998}), and a wider range of radii, known as disk-winds (\citealt{Pudritz1983, Pudritz1986}).

While the jets are associated with material ejected directly from the disk, outflows are described as swept-up material, tracing the interaction between an inner jet or fast stellar winds with the infalling envelope or parent cloud (e.g., \citealt{Shu1991}; \citealt{Matzner1999}; \citealt{Li2013}; \citealt{JALV2019}; \citealt{Shang2023a}). The entrainment of the material can be explained in two different ways, by the jet-driven bow shocks (e.g., \citealt{Masson1993}; \citealt{Raga1993}; \citealt{Lee2001}; \citealt{Ostriker2001}; \citealt{Rabenanahary2022}; \citealt{RiveraOrtiz2023}) or a wide-angle wind (\citealt{Li1996}; \citealt{Ostriker1997}; \citealt{Lee2000}; \citealt{Liang2020}). 
In the first scenario, the jet propagates into the surrounding cloud and forms bow shocks which push and accelerate the ambient gas, producing outflow shells surrounding the jet (e.g., \citealt{Raga1993}). In the second scenario, a radial wind pushes against into the ambient material, forming a radially expanding outflow shell (e.g., \citealt{Lee2000}). These two entrainment processes could act simultaneously, as the jet bow shocks and the wide-angle wind may coexist (e.g., \citealt{Lee2015,Lee2021,Lee2022}). 
A unified model of the outflow is presented in \citet{Shang2006,Shang2020}, where the outflows are formed by the interaction of a fast wide-angle wind launched from the innermost region (consistent with the X-wind scenario) with the ambient medium. 
A unified model of the outflows, where jet-driven and wide-angle scenarios are integrated, is presented in \citet{Shang2006,Shang2020}. They found that the outflows are formed by the interaction of a wind launched from the innermost region (consistent with the X-wind scenario) with the ambient medium. 

Recent radio interferometric and high resolution infrared observations reveal three important characteristics of the outflows. 
First, outflows exhibit internal structures such as nested/multiple shells connecting to the knots and bow shocks driving them (e.g., \citealt{Lee2015}; \citealt{Schutzer2022}), and associated with episodic ejections (e.g., \citealt{Louvet2018}; \citealt{Zhang2019}; \citealt{FernandezLopez2020}; \citealt{Podio2021}; \citealt{dValon2022}; \citealt{Harada2023}; \citealt{Flores2023}; \citealt{JALV2024}), and bubbles (e.g., \citealt{Vazzano2021}). Second, outflows present velocity gradients across the outflow axis, which are interpreted as rotation, e.g., CB 26 (\citealt{Launhardt2009,Launhardt2023}; \citealt{JALV2023}), Ori-S6 (\citealt{Zapata2010}), DG Tau B (\citealt{Zapata2015}; \citealt{dValon2020,dValon2022}), TMC1A (\citealt{Bjerkeli2016a}), Orion Source I (\citealt{Hirota2017}; \citealt{JALV2020}), HH 212 (\citealt{Lee2018,Lee2021}), HH 30 (\citealt{Louvet2018}; \citealt{JALV2024}), NGC 1333 IRAS 4C (\citealt{Zhang2018}), and HD 163296 (\citealt{Booth2021}). 
Finally, the jet-outflow system exhibits an onion-like kinematic and chemical structure, with different layers showing decreasing velocity and associated with the emission of various tracers, such as SO and SO$_2$ in the case of HH 212 (\citealt{Tabone2017}; \citealt{Lee2021}), and CO, H$_2$, and [FeII] in the cases of DG Tau B (\citealt{dValon2020,dValon2022}; \citealt{Delabrosse2024}) and HH 46/47 (\citealt{Nisini2024}).

The protostellar system HH 212 is located in the L1630 cloud of Orion B at a distance of $\sim423\pm21$ pc (\citealt{Zucker2019}). The central source has an age of $\sim40000$ yr (\citealt{Lee2014}) and a mass of $0.25\pm0.05\mathrm{M_\odot}$ (\citealt{Codella2014}; \citealt{Lee2014}). HH 212 is a very well-studied system with a flattened envelope (\citealt{Lee2014,Lee2017ApJ}) accreting material to a protostar-disk system. The system exhibits a dusty edge-on accretion disk (\citealt{Lee2017Sc}) with an inclination angle with respect to the line of sight $i\sim86^{\circ}$ (\citealt{Claussen1998}) and with an atmosphere rich in complex molecules such as CH$_2$DOH, CH$_3$OH, CH$_3$SH, NH$_2$CHO, D$_2$CO, HNCO, H$_2$CO, H$_2$CO$^{13}$, and CH$_3$CHO  (\citealt{Lee2017ApJ,Lee2022mol}). It presents a very prominent SiO rotating jet launched from the innermost disk, $\approx 0.05\, \mathrm{au}$ (\citealt{Podio2015}; \citealt{Lee2017Nat}; \citealt{Codella2023}). In recent studies, \citet{Tabone2017} and \citet{Lee2018,Lee2021}, through the emission of the different transitions of SO and SO$_2$ lines found that the outflow presents signatures of the rotation and an onion-like structure consistent with a radially extended magnetohydrodynamic (MHD) disk wind launched up to 40 au. Additionally, \citet{Lee2021} detected that SO emission exhibits an inner shell at a radius of $\sim$ 4 au, which they proposed could trace the interaction of the extended wide-angle MHD disk wind with large bow shocks driven by the time-variable jet (\citealt{Tabone2018}). Finally, \citet{Lee2022} found that the HH 212 protostellar system has an SiO shell, whose base is associated with the SO inner shell, they showed that the SiO shell shape and kinematics is formed by the radial component of the wide-angle X-wind expanding into the outer disk wind by the cumulative effect of several large jet bow shocks impacting an outer disk wind (\citealt{Tabone2018}).

\begin{table*}[t!]
    \centering
    \caption{Parameters of the observed lines}
    \begin{tabular}{c c c c c}
    \hline
    \hline
    \colhead{Line or Continuum} & \colhead{Rest frequency} & \multicolumn{2}{c}{Synthesized Beam}  & \colhead{Project} \\
    \cline{3-4}
        \colhead{} & \colhead{(GHz)} & \colhead{(arcsec$\times$arcsec)} & \colhead{(deg)}  & \colhead{}  \\
    \hline
    CO ($J=3-2$)  & 345.7959899 & 0.059$\times$0.057 & 51.36  & 2015.1.00024.S \\
       &    &  &   & 2015.1.00041.S \\
       &    &  &   & 2016.1.01475.S \\
    \hline
    SiO ($J=8-7$) & 347.330631  & 0.055$\times$0.050 & 54.53 & 2015.1.00024.S \\
       &    &  &   & 2016.1.01475.S \\
    \hline
    SO ($J=8-7$)  & 346.528481  & 0.057$\times$0.055 & 50.42 & 2015.1.00024.S \\
       &    &   &  & 2016.1.01475.S \\
    \hline
    Continuum     & 358 & 0.020$\times$0.017 & -63.18 & 2015.1.00024.S \\
    \hline
    \end{tabular}
    \tablecomments{The quoted rest frequencies have been extracted from the Molecular Spectroscopy database of the Jet Propulsion Laboratory (JPL) using the online port Splatalogue (\citealt{Remijan2007}) at \url{https://splatalogue.online/\#/home}.}
    \label{tab:obshighres}
\end{table*}

In previous works, \citet{Lee2015} and \citet{Podio2015} reported the emission of the CO molecular line of the HH 212 source at large scales with low angular resolution ($\gtrsim$0$^{\prime\prime}.5$). They observed that the outflow exhibits a nested shell structure and found internal working surfaces associated with the bow shock emission. In this paper, we present high-angular resolution observations of the CO molecular line emission. We focus on a region close to the central source ($z\lesssim\pm1500\,\mathrm{au}$) and we will compare the CO molecular line emission with the previously reported emission of the SiO and SO molecular lines. The paper is organized as follows: in Section \ref{sec:observations} the observations are detailed. We show the results in Section \ref{sec:results}. These results are discussed in Section \ref{sec:discussion}. Finally, our conclusions are presented in Section \ref{sec:conclusions}.

\section{Observations} \label{sec:observations}

In this work, we used three ALMA data sets of the HH 212 protostellar system from the following projects: 2015.1.00024.S (PI: Chin-Fei Lee), 2015.1.00041.S (PI: John Tobin), and 2016.1.01475.S (PI: Claudio Codella). The data were calibrated using the Common Astronomy Software Application (CASA) package (\citealt{CASAteam}) versions 4.5.0, 4.7.2, and 4.7.0 for the projects 2015.1.00024.S, 2015.1.00041.S, and 2016.1.01475.S, respectively. During the imaging process, for all data sets, a Briggs weighting robust parameter of 0.5 was applied to balance sensitivity and angular resolution. 

Band 7 high angular resolution of the HH 212 protostellar system was carried out with ALMA using 32-45 12 m antennas with two executions in 2015, one on November 5 and the other on December 3, during the Early Science Cycle 3 phase (2015.1.00024.S), 34-42 12 m antennas on 2016 September 6 and 2017 July 19 during the Cycle 4 phase (2015.1.00041.S), and 44 12 m antennas between October 6 and November 26, 2016 during the Cycle 4 phase (2016.1.01475.S). The maximum baselines are 16196 m, 3697 m, and 3000 m for the Cycles 3 and 4, respectively. The details of the observations have been reported in \citet{Lee2017ApJ} for Cycle 3, \citet{vanthoff2018} and \citet{Tobin2019} for Cycle 4 (2015.1.00041.S), and \citet{Bianchi2017} and \citet{Tabone2017} for Cycle 4 (2016.1.01475.S). 

The continuum image rms noise is 0.15 mJy/Beam and the line images have an rms noise level of 2.11 mJy/Beam, 0.96 mJy/Beam, and 0.92 mJy/Beam for the CO, SiO, and SO, respectively. Details of the synthesized beam and rest frequencies are provided in Table \ref{tab:obshighres}.

\begin{figure*}[t!]
\centering
\includegraphics[width=\linewidth]{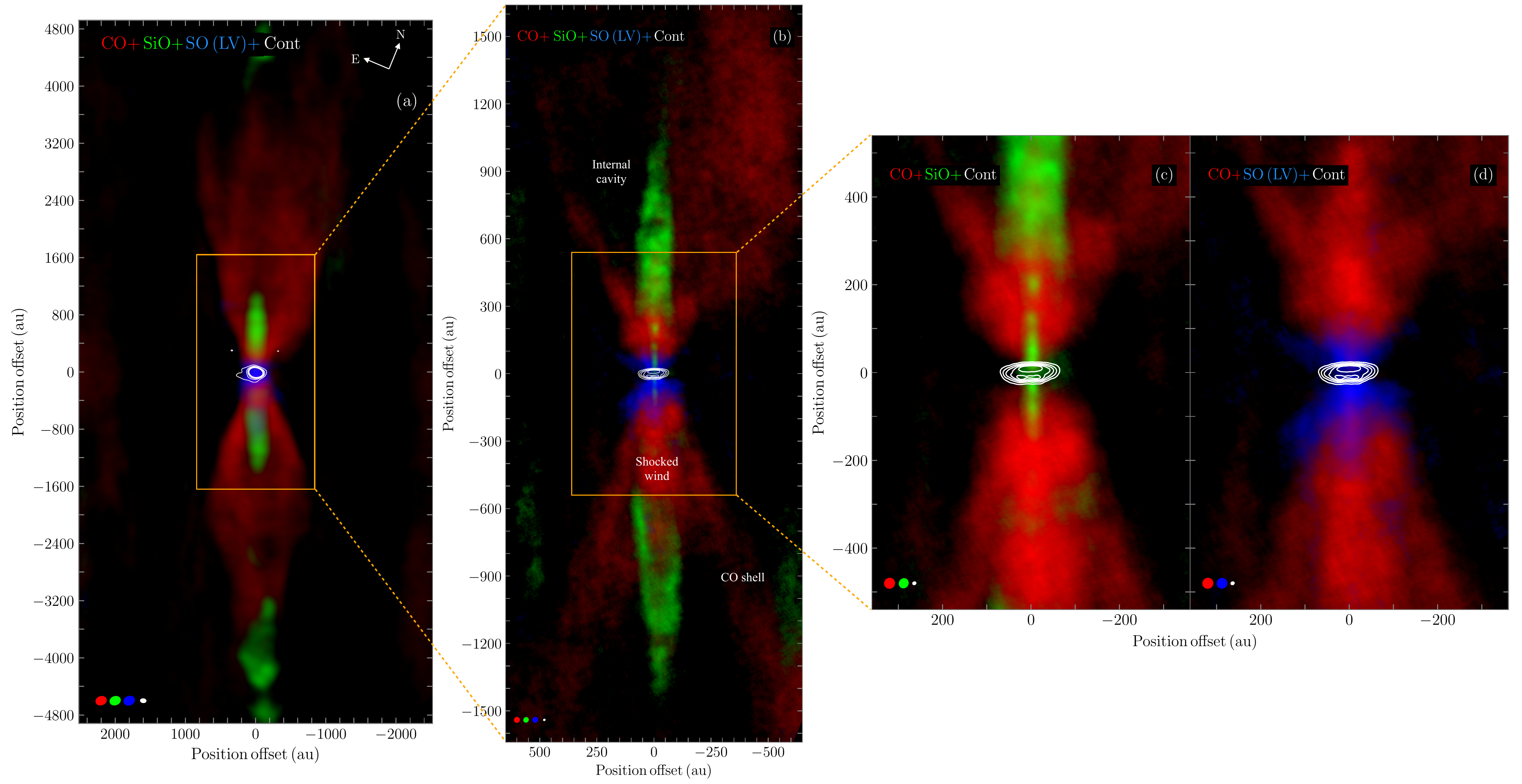}
\caption{Composite images of the HH 212 protostellar system of the ALMA moment zero or integrated intensity of CO, SiO, and SO low-velocity (LV) lines together with the 358 GHz continuum map of the disk. (a) All lines and continuum obtained with low angular resolution data, $\sim$0$^{\prime\prime}$.37. (b) A zoomed-in image to the center of the system of all molecules and continuum obtained with high angular resolution data, $\sim$0$^{\prime\prime}$.05. (c) A zoomed-in image of the center of the system of the CO and SiO molecular line emission, together with a continuum map. (d) A zoomed-in image of the center of the system of the CO and SO (LV) molecular line emission, together with a continuum map. 
Color codes are consistent across all labels. White contours start from 3$\sigma$ in steps of 3$\sigma$, 6$\sigma$, 9$\sigma$, and 12$\sigma$, where $\sigma= 1.24$ mJy/Beam for low angular resolution data and $\sigma=0.15$ mJy/Beam for high angular resolution data. The synthesized beams are shown in the lower left corner of all panels.}
\label{fig:molecules_resolution}
\end{figure*}

\section{Results} \label{sec:results}

To provide readers with a more comprehensive view of the HH 212 system, we used the low angular resolution dataset ($\sim$0$^{\prime\prime}$.37) from the following projects: 2011.0.00647.S (PI: Claudio Codella), 2012.1.00122.S (PI: Chin-Fei Lee), and 2016.1.01475.S (PI: Claudio Codella). These datasets help contextualize the different scales for our analysis. Additionally, the continuum maps are used as a reference to locate the central object. This paper is focused on presenting and discussing the emission of the CO molecular line; however, the SiO and SO molecular lines are also presented here for our comparison. Also, for clarity in the figures, all the maps are rotated by $22^\circ.5$ clockwise so that the jet axis aligns with the $y-$axis.

The systemic velocity in HH 212 is $V_\mathrm{sys}=1.7\pm0.1\,\mathrm{km\,s^{-1}}$ (\citealt{Lee2014}). To facilitate the interpretation of the results, we define an offset velocity $V_\mathrm{off}= V_\mathrm{lsr}-V_\mathrm{sys}$.

\begin{figure*}[t!]
\centering
\includegraphics[width=\linewidth]{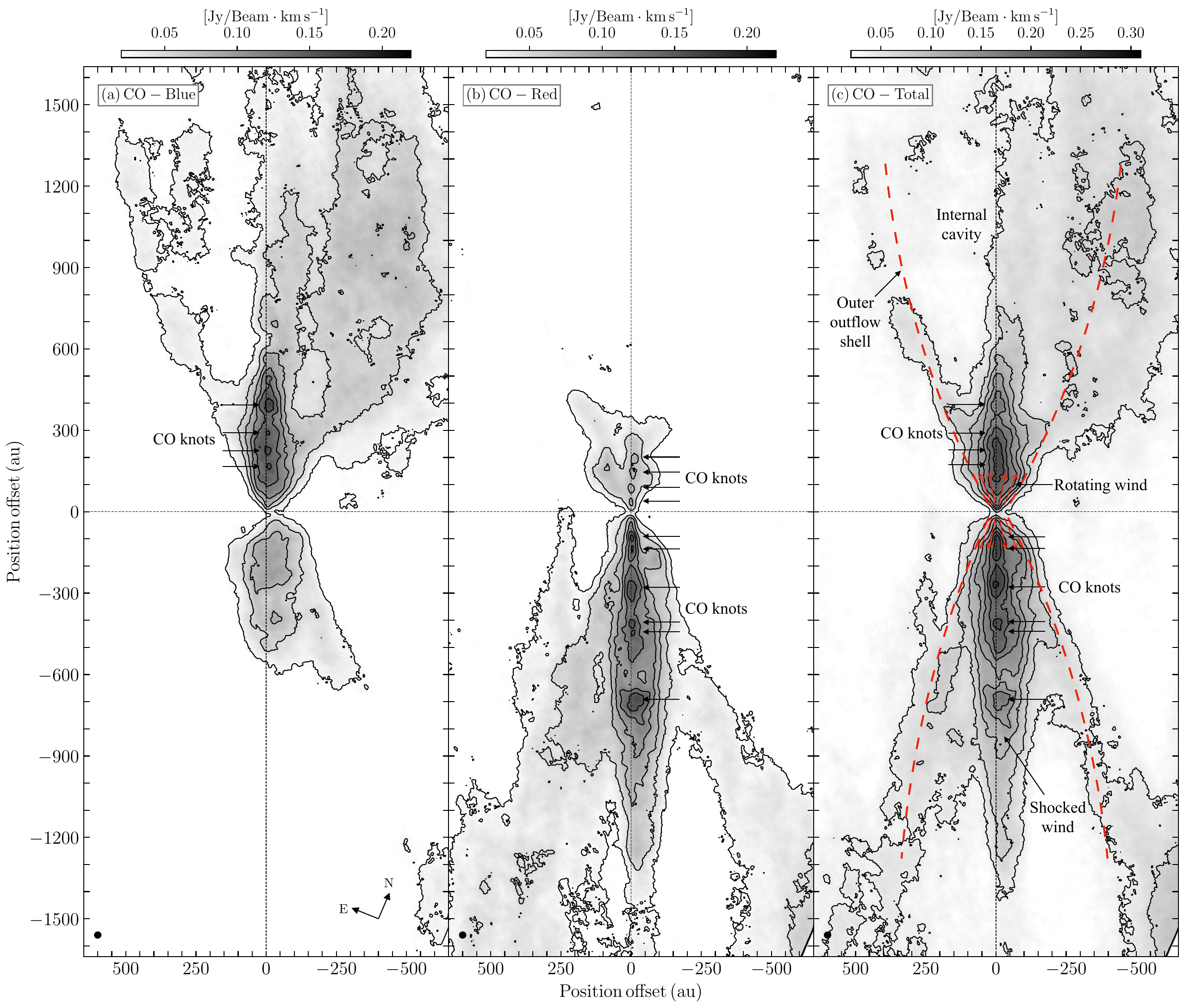}
\caption{The CO maps of the outflow associated with HH 212. (a) Blue-shifted emission integrated from 1.7 to 12.7 $\mathrm{km\,s^{-1}}$. (b) Red-shifted emission integrated from -9.3 to 1.7 $\mathrm{km\,s^{-1}}$. (c) Full range of the emission $\pm 11\,\mathrm{km\,s^{-1}}$ around the systemic velocity. The contour levels start from 3$\sigma$ in steps of 3$\sigma$, 6$\sigma$, 9$\sigma$, 12$\sigma$, and 15$\sigma$, where $\sigma=8.48$ mJy/Beam$\cdot\mathrm{km\,s^{-1}}$, $\sigma=8.01$ mJy/Beam$\cdot \mathrm{km\,s^{-1}}$ for panels (a) and (b), respectively. While, in panel (c) the contour levels start from 3$\sigma$ in steps of 2$\sigma$, 4$\sigma$, 6$\sigma$, 8$\sigma$, 10$\sigma$, 12$\sigma$, and 14$\sigma$, where $\sigma=13.92$ mJy/Beam$\cdot\mathrm{km\,s^{-1}}$. The black arrows mark the position of the CO knots and the shocked wind, while the other components of the protostellar system such as CO shell and wind are depicted by the red dashed lines and red arrows, respectively. The synthesized beams are shown in the lower left corner of all panels.}
\label{fig:moment0}
\end{figure*}

\subsection{Comparison between Molecules}
\label{subsec:comparison}

Figure \ref{fig:molecules_resolution} shows the ALMA moment zero or integrated intensity maps of the emission of the CO (red), SiO (green), and SO low-velocity (LV) (blue) molecular lines overlaid on the white contours of the 358 GHz continuum map of the disk (adopted from \citealt{Lee2017Sc}) in the protostellar system HH 212. The CO and SiO maps were integrated over a velocity range of $\pm11\,\mathrm{km\,s^{-1}}$, while the SO (LV) was integrated over a range of $\pm 3\,\mathrm{km\,s^{-1}}$; all velocities are relative to the systemic velocity $V_\mathrm{sys}$. For the SO emission, we use the low-velocity emission that traces the outflow, as shown in \citet{Lee2021}. Another reason that we do not consider the high-velocity emission of the SO is because this line falls outside of the spectral set-up of the dataset (\citealt{Tabone2017}). 

Figure \ref{fig:molecules_resolution}a displays the map obtained with the low angular resolution data of the outflow presented by \citet{Lee2015}, within $\pm4800\,\mathrm{au}$ of the central star. The high angular resolution data are presented in Figure \ref{fig:molecules_resolution}b, where the emission is plotted within $\pm1600\,\mathrm{au}$ of the central star. In both figures, the CO emission reveals an outflow shell previously reported by \citet{Lee2015}. Additionally, both CO and SO trace the molecular outflow near to the accretion disk ($z\lesssim |200|\,\mathrm{au}$). The jet and a shell around the jet are traced by the SiO emission (see \citealt{Lee2017Nat,Lee2022}).  The high angular resolution data suggest that CO is likely tracing the shocked wind within the SiO shell. Notably, while CO, SiO, and SO emissions coexist in the central region, the external region is predominantly dominated by CO emission.

\begin{figure}[t!]
\centering
\includegraphics[scale=0.6]{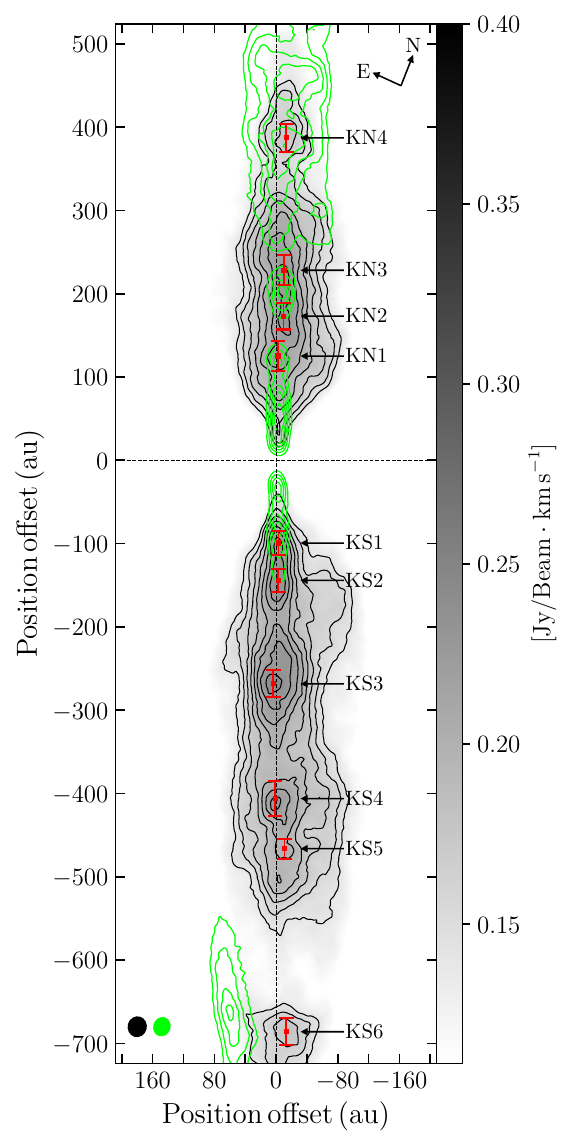}
\caption{Zoom around the jet of the CO maps of the HH 212 protostellar system of the CO molecular line emission. Green contours represent ALMA moment zero of the SiO molecular line emission. Red points show the position of the CO knots. The contours labels start from 10$\sigma$ in steps of 1$\sigma$, 2$\sigma$, 3$\sigma$, 4$\sigma$, 5$\sigma$, 6$\sigma$, 7$\sigma$, and  8$\sigma$, where $\sigma=13.92$ mJy/Beam$\cdot\mathrm{km\,s^{-1}}$ for the CO emission and start from 9$\sigma$ in steps of 2$\sigma$, 4$\sigma$, 6$\sigma$, and 8$\sigma$, where $\sigma=9.57$ mJy/Beam$\cdot\mathrm{km\,s^{-1}}$ for the SiO emission. The synthesized beams are shown in the lower left corner.}
\label{fig:zoom_knots}
\end{figure}

\begin{figure}[t!]
\centering
\includegraphics[width=\linewidth]{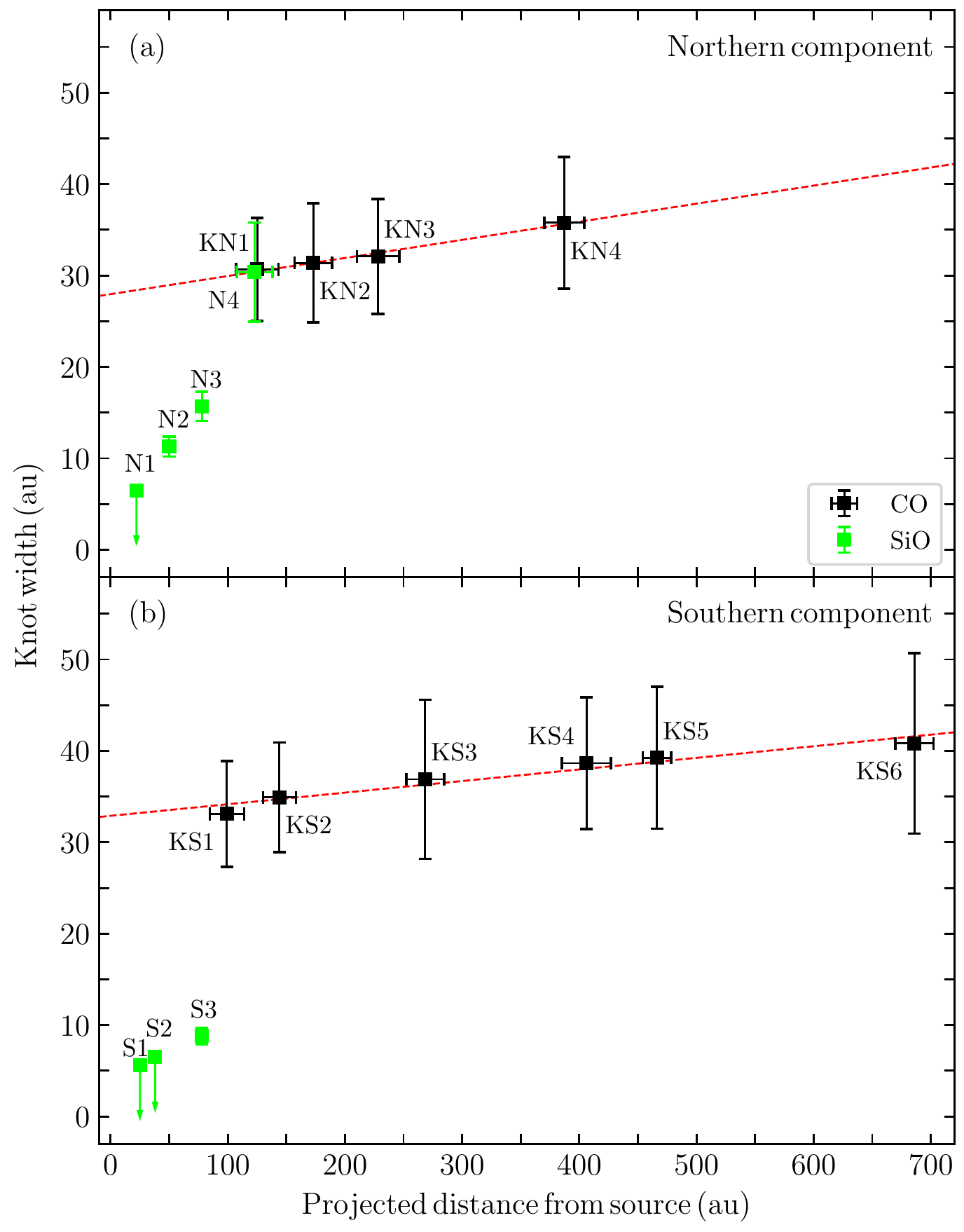}
\caption{The width of the CO knots position as a function of the projected distance to the central source. (a) Detected knots in the northern component. (b) Detected knots in the southern component. Green points depicted the SiO knots taken from \citet{Lee2017Nat}. The error bars show the uncertainties in the width and the position.}
\label{fig:knots_position}
\end{figure}

The comparison between CO molecular line emission and the SiO jet, as well as the SO (LV) molecular outflow within $\pm 500\,\mathrm{au}$ is presented in Figures \ref{fig:molecules_resolution}c and d, respectively. Figure \ref{fig:molecules_resolution}c illustrates the comparison between CO outflow and the SiO jet. At low heights ($z\lesssim |200|\,\mathrm{au}$), the jet is dominated by knots, while CO traces the outflow around the jet.
On the other hand, the comparison between the CO and SO (LV) molecular outflows are presented in Figure \ref{fig:molecules_resolution}d. We observe that the wind previously detected in SO and identified as a wide-angle disk wind (\citealt{Tabone2017}; \citealt{Lee2021}) coexists with the CO\@. It is important to note that in a region around the central star ($z\lesssim |600|\,\mathrm{au}$), the HH 212 protostellar outflow exhibits an onion-like structure. 

In Figure \ref{fig:molecules_resolution}, the outermost region relative to the jet axis is dominated by the emission of the CO outer shell, which connects with the knots and bow shocks driven by them observed along the jet axis at large scales, as previously reported by \citet{Lee2015}. At heights of $z\geq |400|\,\mathrm{au}$, we detect an internal cavity between the CO outer shell and the SiO shell. For heights $z\leq |400| \,\mathrm{au}$, outside the SiO shell, the emission is dominated by the SO and CO wind. Finally, inside the SiO shell, we observe the CO wind and the SiO jet.

\subsection{CO Structure: an Outer Outflow Shell, Internal Cavity, Rotating Wind, Shocked Wind, and Jet}
\label{subsec:structure}

\begin{figure*}[t!]
\centering
\includegraphics[width=\linewidth]{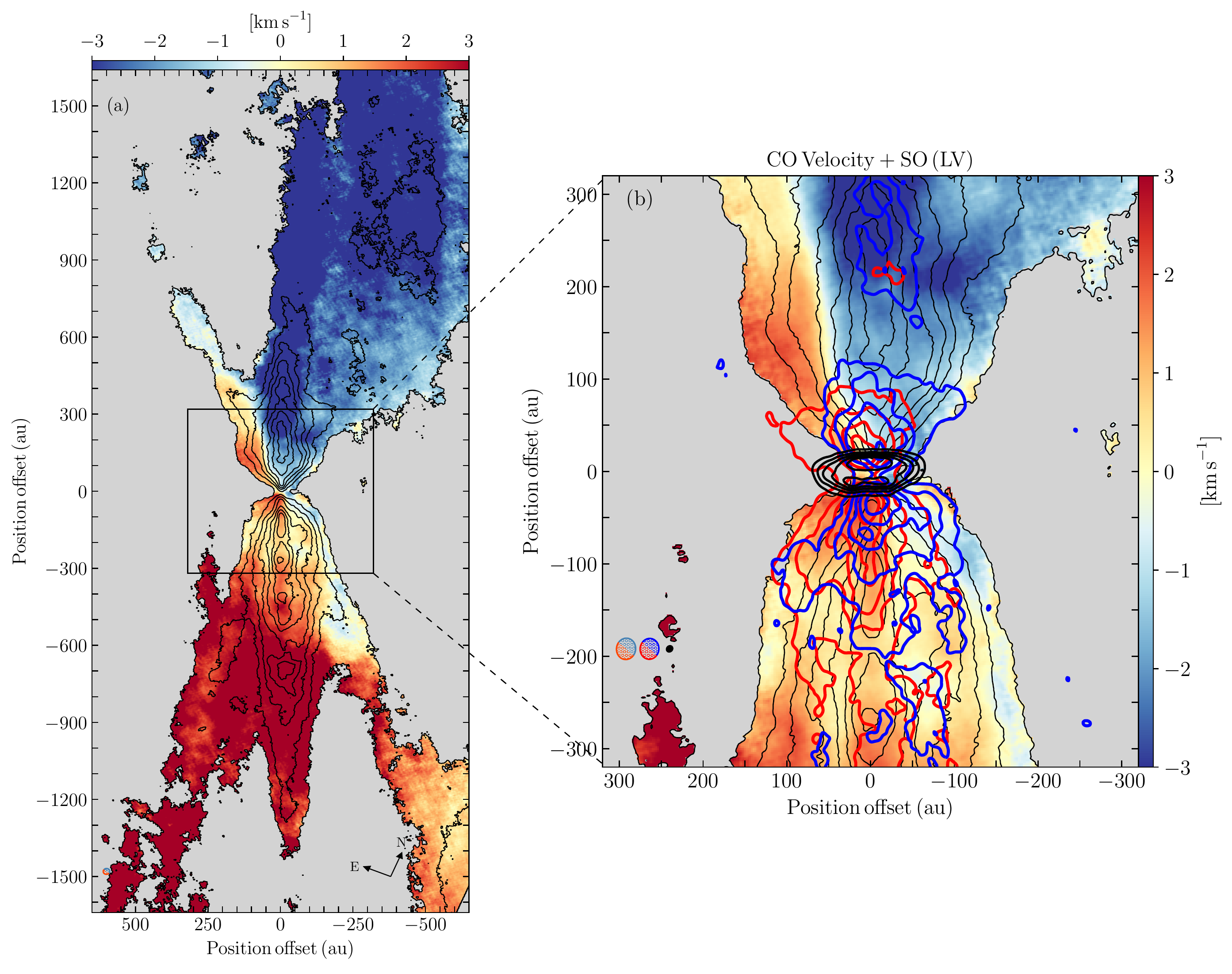}
\caption{ALMA first moment or intensity-weighted velocity of the CO molecular line emission overlaid in black contours the CO total intensity map. (a) Map of the extended emission. (b) Zoom in the central region. Red and blue contours represent the blueshifted and redshifted, respectively, of the intensity map of SO (LV) molecular line emission adopted from \citet{Lee2021}, while the black contours depicted the 358 GHz continuum map of the disk adopted from \citet{Lee2017Sc}. Black contour levels start from 3$\sigma$ in steps of 3$\sigma$ in steps of 2$\sigma$, 4$\sigma$, 6$\sigma$, 8$\sigma$, 10$\sigma$, 12$\sigma$, and 14$\sigma$, where $\sigma=13.92$ mJy/Beam$\cdot \mathrm{km\,s^{-1}}$. The countours of the continuum emission contours start from 3$\sigma$ in steps of 3$\sigma$, 6$\sigma$, 9$\sigma$, and 12$\sigma$, where $\sigma=$ 1.24 mJy/Beam. The synthesized beams are shown in the lower left corner of both panels.}
\label{fig:moment1}
\end{figure*}

The CO maps integrated over different velocity ranges are shown in Figure \ref{fig:moment0}. Figures \ref{fig:moment0}a and \ref{fig:moment0}b display the blueshifted and redshifted CO emission, respectively, integrated over velocities ranging from -9.3 to 1.7 $\mathrm{km\,s^{-1}}$ and from 1.7 to 12.7 $\mathrm{km\,s^{-1}}$, relative to the systemic velocity $V_\mathrm{sys}$. The northern component of the outflow is traced by the blueshifted emission, while the southern component dominates the redshifted emission. However, within $z\lesssim |400|\,\mathrm{au}$, the outflow emits in both velocity regimes. Figure \ref{fig:moment0}c displays the total CO intensity map integrated over all velocity ranges, $\pm11\,\mathrm{km\,s^{-1}}$ away from the systemic velocity $V_\mathrm{sys}$. The innermost contours in Figure \ref{fig:moment0} trace possible CO knots associated with the jet. Additionally, Figure \ref{fig:moment0}c shows an  outer outflow shell (red dashed lines), an internal cavity, and emission features around $z\lesssim |900|\,\mathrm{au}$ including the rotating wind (red arrows) and CO shocked wind inside the SiO shell (see Figure \ref{fig:molecules_resolution}). We infer that the CO shell is produced by the interaction of ejected material from the protostar-disk system with the envelope. 
While the region between the CO and SiO shells may trace the material directly ejected from the disk, so-called disk wind and previously reported by the SO and SO$_2$ emission by \citet{Tabone2017} and \citet{Lee2018,Lee2021}, or it could trace the shocked wind resulting from the interaction of the wind launched directly from the innermost disk with the ambient medium, known as shocked wind (\citealt{Shang2023a}). 
For the reader's convenience, and considering that this region has been extensively studied in previous works and the gas presents signatures of the rotation (e.g., \citealt{Tabone2017,Lee2018,Lee2021}) and shocks, we will henceforth refer to this region as the so-called ``rotating wind'', regardless of its origin. 
Finally, the CO emission inside the SiO shell may trace the shocked wind resulting from the interaction between the wide-angle component of the X-wind with the surroundings (envelope or rotating wind) or it could trace the material ejected sideways from the knots and bow shocks along the jet axis (e.g., \citealt{Tabone2018,Rabenanahary2022,RiveraOrtiz2023}). For simplicity, and since the CO emission traces the interaction of different components of the protostellar system, we will refer to this region as the ``shocked wind''.

\begin{figure*}[t!]
\centering
\includegraphics[scale=0.505]{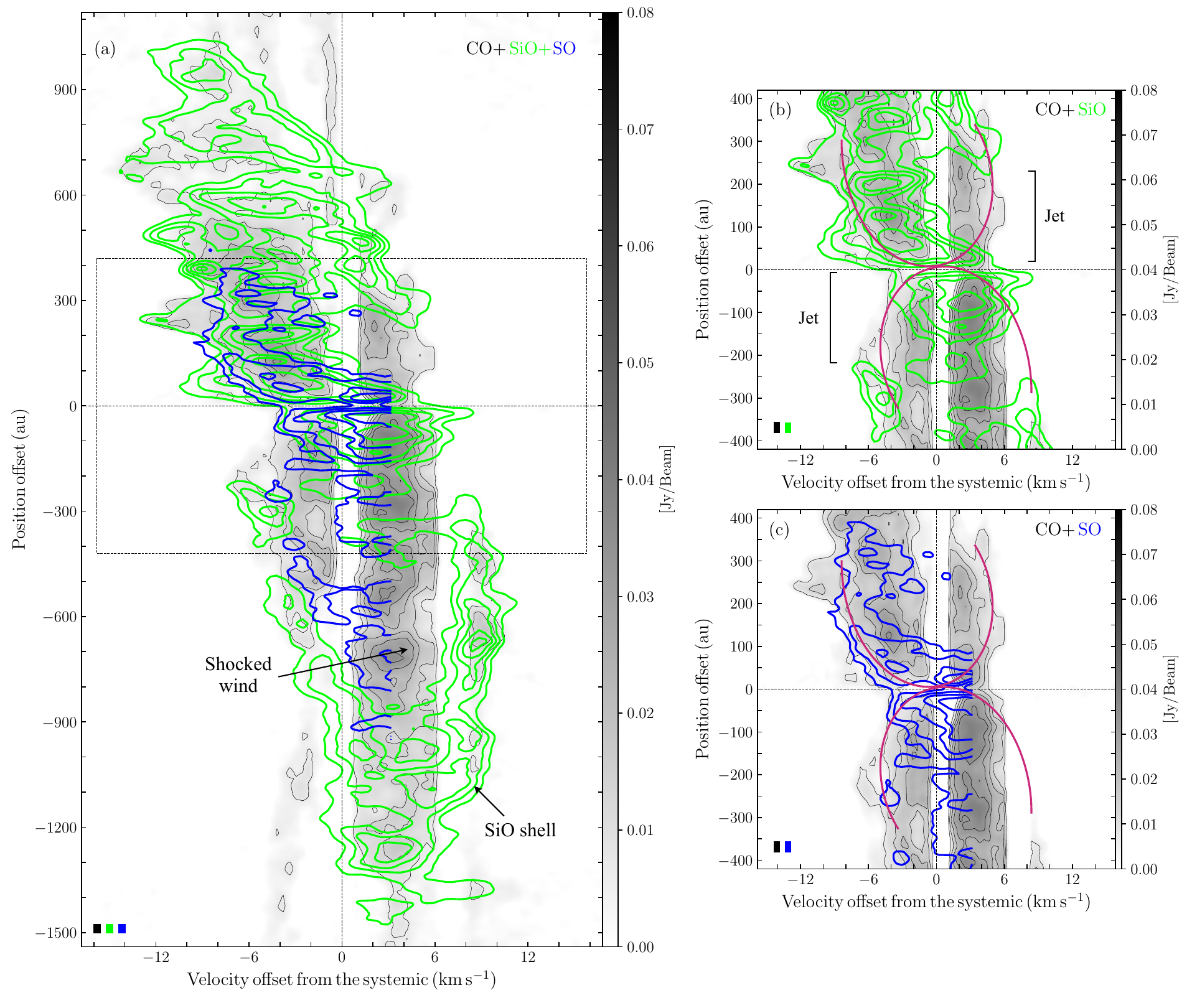}
\caption{The position-velocity diagram along the jet axis of the CO, SiO, and SO molecular line emission. (a) Comparison between the three molecular lines. (b) Comparison between CO and SiO emission. (c) Comparison between CO and SO emission. The color codes for the labels remain consistent across all panels. Magenta curves are included to guide readers for the parabolic PV structure. In all panels the contours start from 3$\sigma$ in steps of 3$\sigma$, 6$\sigma$, 9$\sigma$, and 12$\sigma$ where $\sigma=2.11$ mJy/Beam$\cdot\mathrm{km\,s^{-1}}$, $\sigma=0.96$ mJy/Beam$\cdot\mathrm{km\,s^{-1}}$, $\sigma=0.92$ mJy/Beam$\cdot\mathrm{km\,s^{-1}}$ for the CO, SiO, and SO, respectively. The bars in the left lower corner represent of each panel represent the angular resolution and the channel width used for the PV cut.}
\label{fig:pv_jetaxis}
\end{figure*}

To provide a more detailed view of the chain of CO knots associated with the jet, Figure \ref{fig:zoom_knots} presents a zoomed-in view of the CO total intensity map around the jet axis, overlaid with the SiO map (green contours). Only the emissions exceeding 10$\sigma$ for CO and 9$\sigma$ for SiO are plotted to better illustrate their distinct distributions. Red points mark the positions of the CO knots in the northern (KN1--KN4) and southern (KS1--KS6) components. The positions of the knots were determined from the emission peaks with detections exceeding $>10\sigma$, as identified from the Gaussian fits. In contrast to the chain of SiO knots (\citealt{Lee2017Nat}), the CO knots are detected further away from the central star ($z\gtrsim |100|\,\mathrm{au}$).

\begin{figure*}[t!] 
\centering
\includegraphics[scale=0.55]{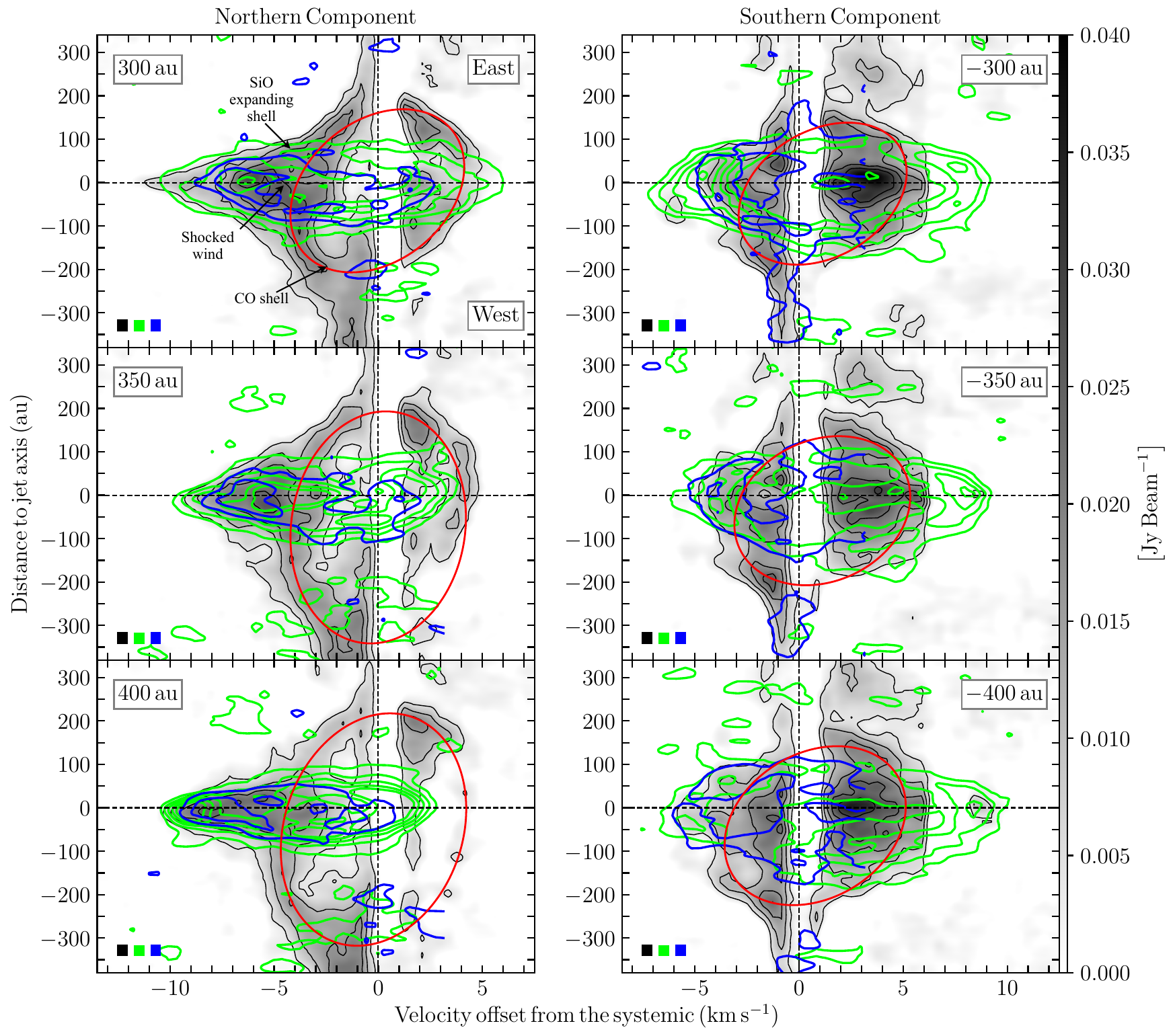}
\caption{Position-velocity diagrams perpendicular to the jet axis of the molecular line emission from the CO (grayscale), SiO (green), and SO (blue) are presented at various heights from $z=\pm300\,\mathrm{au}$ to $z=\pm400\,\mathrm{au}$ with an interval of $50\,\mathrm{au}$. The black, green, and blue bars represent the angular resolution and the channel width. The contour labels start at 2$\sigma$ in steps of 3$\sigma$, 6$\sigma$, 9$\sigma$, and 12$\sigma$, where $\sigma=2.11$ mJy/Beam$\cdot\mathrm{km\,s^{-1}}$, $\sigma=0.96$ mJy/Beam$\cdot\mathrm{km\,s^{-1}}$, $\sigma=0.92$ mJy/Beam$\cdot\mathrm{km\,s^{-1}}$ for the CO, SiO, and SO, respectively. The horizontal dashed line represents the position of the jet axis, and the vertical dashed line indicates $V_\mathrm{off}$. The red ellipses in all panels represent the best fit of the emission of the bright outer CO shell.}
\label{fig:pv_xwinds}
\end{figure*}

\begin{figure*}[t!] 
\centering
\includegraphics[scale=0.585]{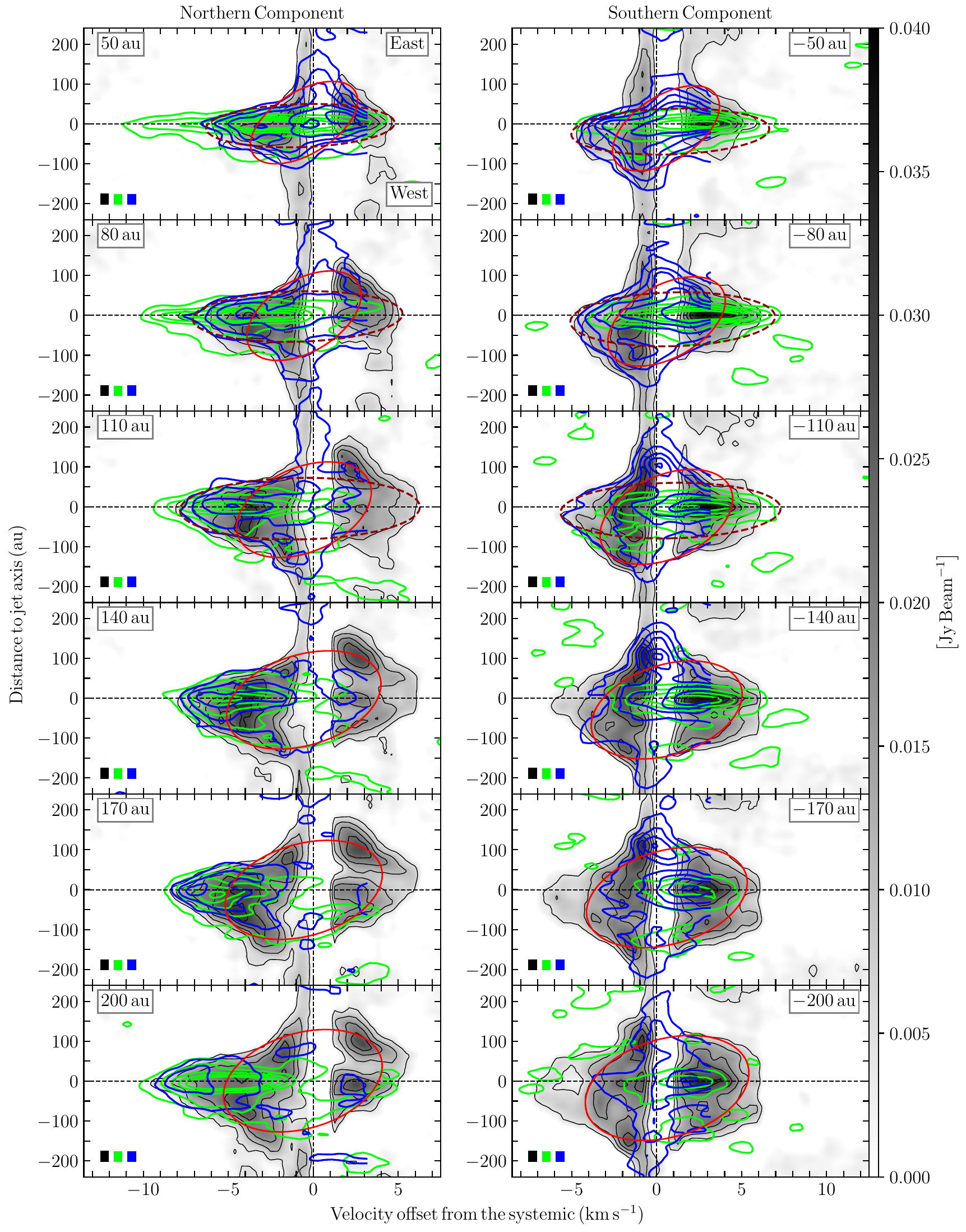}
\caption{Position-velocity diagrams perpendicular to the jet axis of the molecular line emission from the CO (gray scale), SiO (green), and SO (blue) at different heights from $z=\pm50\,\mathrm{au}$ to $z=\pm200\,\mathrm{au}$ with an interval of $30\,\mathrm{au}$. The ellipses in all panels represent the best fit for the SO and CO emission limit at $2\sigma$. The red ellipses exclude the high-velocity components, and the dashed brown ellipses trace the high-velocity portion of the data. 
The description is the same as Figure \ref{fig:pv_xwinds}.}
\label{fig:pv_diskwinds}
\end{figure*}

Figure \ref{fig:knots_position} shows the width of the CO knots as a function of the projected distance from the central source. To determine the width of the knots, we used the method developed in \citet{Lee2017Nat}. We extracted spatial profiles from the CO intensity map (Figure \ref{fig:zoom_knots}) and fitted these profiles with a Gaussian profile. The plot reveals a slight increase in the width of the knots with the projected distance, as indicated by the dashed red lines in both panels. The positions and widths of the SiO knots in the northern (N1-N3) and southern (S1-S3) components were taken from \citet{Lee2017Nat}, with the exception of knot N4. The position and width of knot N4 were measured using the same method applied to the CO knots.

Figures \ref{fig:zoom_knots} and \ref{fig:knots_position} show that the CO knots are associated with four different episodic ejections: KN1--KN3 (together with KS1--KS2), KN3 (with KS3), KN4 (with KS4--KS5), and KS6. The average distance between these ejections is $140\pm14\,\mathrm{au}$, with a timescale of $5.2\pm0.6$ yr, assuming a jet velocity of $\sim135\,\mathrm{km\,s^{-1}}$ (\citealt{Zinneckar1998,Lee2017Nat}). Within these episodic ejections, we observe that the knots (KN1--KN2--KN3, KS1--KS2, and KS4--KS5) are separated by an average distance of $49\pm13\,\mathrm{au}$, with a timescale of $1.8\pm0.4$ yr. This suggests that the jet exhibits quasi-periodicity. This quasi-periodicity may be associated with variations in the ejection velocity of the jet (\citealt{Moraghan2016}) or with magnetic pseudopulses, which tends to be periodic and can create perturbations in density and velocity, resulting in the appearance of knots (\citealt{Shang2002,Shang2020}).

\subsection{Kinematic Properties of the Protostellar System: Rotating Outflow Shells and Interaction between Wide-Angle Winds and the Ambient Envelope}
\label{subsec:kinematical}

Figure \ref{fig:moment1} displays the first moment, or intensity-weighted velocity map of the CO emission line, overlaid by the black contours of the CO total intensity map. In Figure \ref{fig:moment1}a, the emission is shown up to $\pm 1600\,\mathrm{au}$, with blueshifted velocities observed on the east-side and redshifted velocities on the west-side. Figure \ref{fig:moment1}b focuses on the central region, with red and blue contours indicating the redshifted and blueshifted low-velocity emission from the SO molecular line, respectively, and the black contours show the 358 GHz continuum map of the disk (adopted from \citealt{Lee2017Sc}). It is noteworthy that the outflow exhibits a velocity gradient across the outflow axis, which is associated with the rotation of gas at $\sim1.5\,\mathrm{km\,s^{-1}}$. This rotation velocity is consistent with previous reports of the emission from SO and SO$_2$ by \citet{Tabone2017} and \citet{Lee2018,Lee2021} and it has the same direction as the rotation of the disk, envelope, and SiO jet (e.g., \citealt{Codella2014,Lee2014,Lee2017Nat}). Furthermore, further away from the central source, the outflow is predominantly blueshifted in the north and redshifted in the south, resembling the velocity sense of the bow shocks at larger distances, thus indicating that they are driven by them (\citealt{Lee2015}), and showing strong dominance and influence from the surrounding envelope and the complex interactions weighted by uneven density inhomogeneity on both the blue side but not on the red side. Additionally, in Figure \ref{fig:moment1}b, the red SO emission clearly extends outside of the CO, which may indicate an extended and deeper connection to the outer envelope or it could simply be due to CO absorption by the cloud at $V\sim1.7\,\mathrm{km\,s^{-1}}$, at seen in Figure \ref{fig:pv_diskwinds}.

The position-velocity (PV) diagram along the jet axis presented in Figure \ref{fig:pv_jetaxis} traces the emission of the CO (gray scale), SiO (green contours), and SO (blue contours) molecular lines. In Figure \ref{fig:pv_jetaxis}a, we find that further away from the central source, the SiO emission traces a shell  in the PV (\citealt{Lee2022}), while the CO and SO emission is dominated by the shocked wind. Close to the central source (dashed rectangle in Figure \ref{fig:pv_jetaxis}a, $z\leq|400|\,\mathrm{au}$), the SiO emission (Figure \ref{fig:pv_jetaxis}b) traces the jet, while the SO emission (Figure \ref{fig:pv_jetaxis}c) in the northern region exhibits a classic parabola shape (magenta lines) previously reported by \citet{Lee2018}, in the southern region, we do not detect this shape as the SO falls outside of the spectral set-up (\citealt{Tabone2017}). This parabola is associated with an expanding shell swept up by successive bowshocks (\citealt{Lee2021}) or by wide-angle X-wind (\citealt{Lee2022}).

Considering that the CO molecular line emission of the HH 212 protostellar system traces various components, such as an outer outflow shell, the rotating wind, the shocked wind, and a chain of the knots associated with the jet, we will separate the observed region by height into two different regions, further away and close to the central source, $z>|200|\,\mathrm{au}$ and $z\leq |200|\,\mathrm{au}$, respectively. The lack of CO emission around $V_\mathrm{off}=0$ is because the outflow emission at this velocity is absorbed by the parent cloud. Similarly, the lack of CO emission around $V_\mathrm{off}\sim7\,\mathrm{km\,s^{-1}}$ is because the outflow emission at this velocity is absorbed by another cloud in front of HH 212 \citep{Lee2000}.

\subsubsection{Kinematic Properties Further at Larger than 200 au}
\label{subsubsec:kin_far}

As observed in Figure \ref{fig:pv_jetaxis}, the source presents a structure of multiple shells with different velocities, also known as an onion-like velocity structure. To confirm this, PV diagrams perpendicular to the jet axis are displayed in Figure \ref{fig:pv_xwinds}. The PV diagrams were made from $z=\pm300\,\mathrm{au}$ to $z=\pm400\,\mathrm{au}$ every 50 au. The CO emission is shown in grayscale and the SiO and SO emissions are plotted in green and blue contours, respectively. In all panels, we observe that the CO traces two different regions, a low-velocity extended shell (red ellipses) associated with the  outer outflow shell presented in Figures \ref{fig:molecules_resolution} and \ref{fig:moment0}, and a high-velocity component around the jet axis, which may indicate the presence of a shocked wind. The SiO molecular line emission forms a shell between the emission of the CO high-velocity component and the low-velocity extended shell, while the SO emission may be tracing the region between the SiO shell and the CO high-velocity component. 

\begin{figure*}[t!]
 \centering
\includegraphics[scale=0.65]{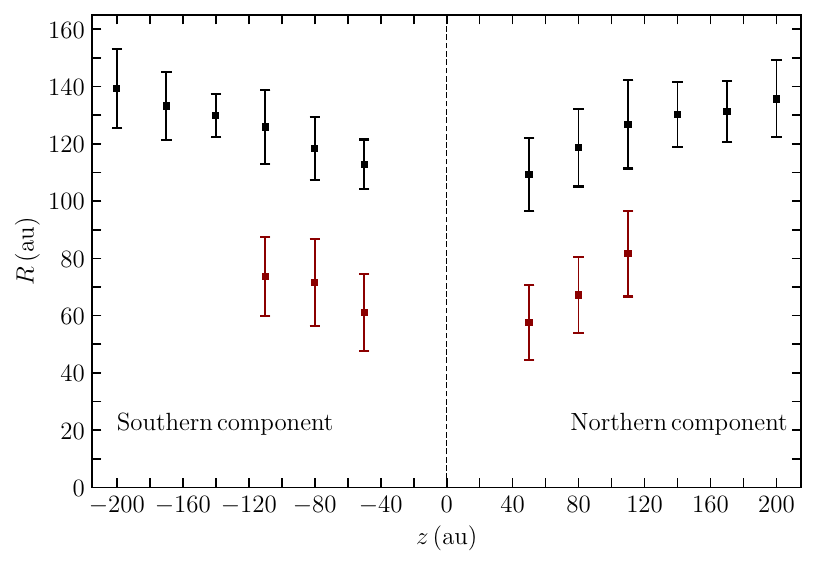}\includegraphics[scale=0.65]{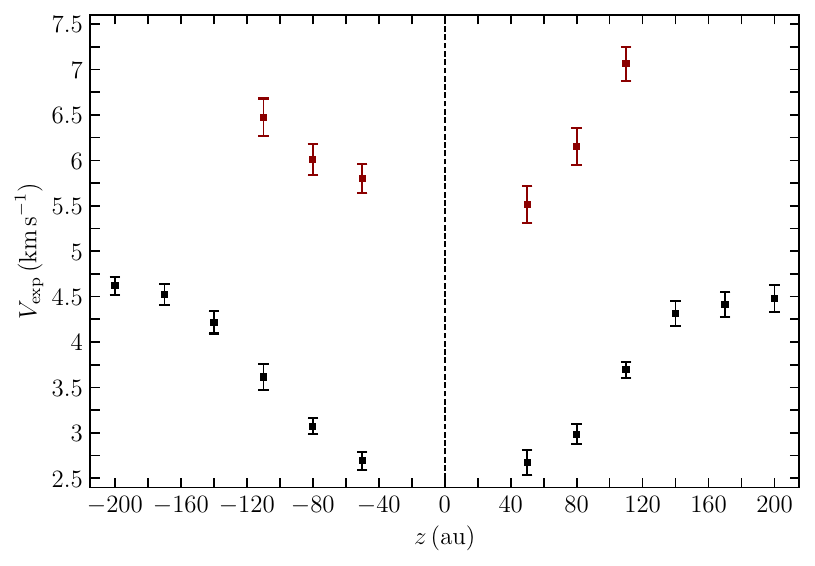}
\includegraphics[scale=0.65]{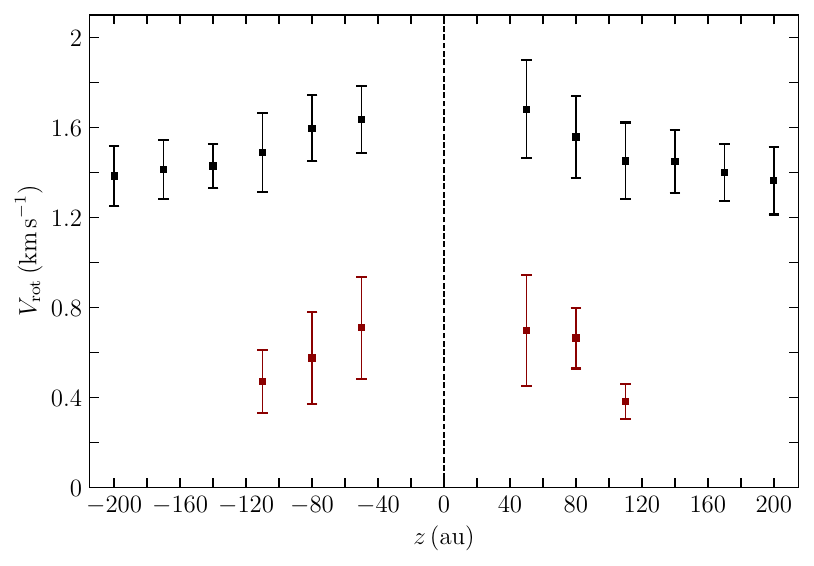}\includegraphics[scale=0.65]{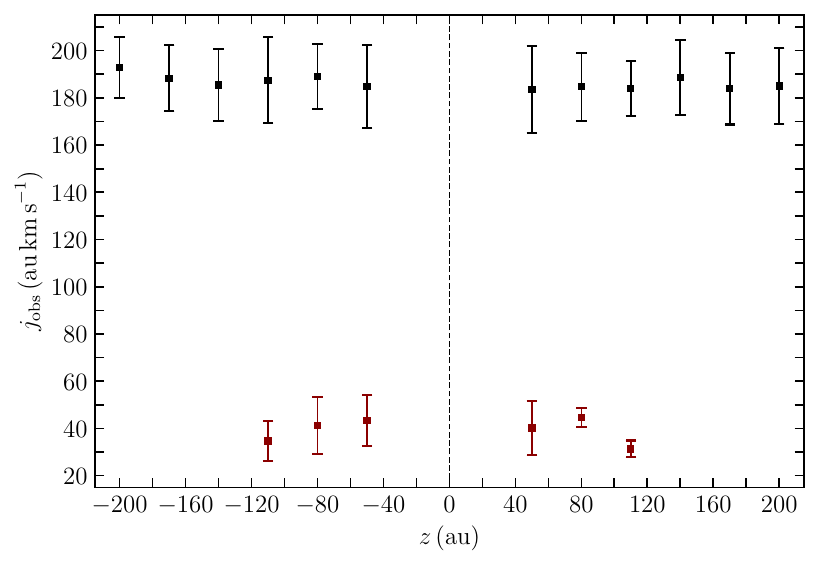}
\caption{The kinematical properties of the CO outflow of HH 212 as a function of the height with respect to the disk mid-plane $z=0$. Top left: Cylindrical radius $R$. Top right: Radial expansion velocity $V_\mathrm{exp}$. Bottom left: Rotation velocity $V_\mathrm{rot}$. Bottom right: Specific angular momentum $j_\mathrm{obs}$. The error bars represent the uncertainties in the measurements of these kinematical properties. The brown points with error bars are derived from those dashed ellipses in Figure \ref{fig:pv_diskwinds}, for the three different heights at 50, 80, and 110 au from the midplane.}
\label{fig:dwparameters}
\end{figure*}

\subsubsection{Kinematic Properties Close to Central Source}
\label{subsubsec:kin_close}

PV diagrams perpendicular to the jet axis at different heights close to the central source ($z\leq |200|\,\mathrm{au}$) are shown in Figure \ref{fig:pv_diskwinds}. The diagrams were made from $z=\pm 50\,\mathrm{au}$ to $z=\pm200\,\mathrm{au}$ every 30 au. The CO emission is traced in grayscale, while the emission of the SiO and SO is plotted in green and blue contours, respectively. At heights between $|50|\,\mathrm{au}\leq z\leq |110|\,\mathrm{au}$, the CO and SO emissions are spatially consistent and exhibit the same velocity, while the SiO emission is dominated by the jet and traces a narrow region with a high velocity dispersion. In contrast, at heights $z>|110|\,\mathrm{au}$, the CO and SO coexist spatially; however, the SO traces gas of low velocity and the CO traces gas over a wider range of the velocities with respect to SO,  lower than those of the SiO\@. The SiO shows regions close to the jet axis, but with less velocity dispersion than in the regions closer to the disk.

\begin{table*}[t!]
    \centering
    \caption{Parameters and kinematical properties obtained from the elliptical fits in the PV diagrams across the outflow axis}
    \begin{tabular}{c c c c c c c c c c c}
    \hline
    \hline
    \colhead{Component} & \colhead{$z$} & \colhead{$r_\mathrm{cent}$} & \colhead{$|V_\mathrm{cent}-V_\mathrm{sys}|$} & \colhead{$a$} & \colhead{$b$} & PA & $R$ & $V_\mathrm{exp}$ & $V_\mathrm{rot}$ & $V_{z}$ \\
    \colhead{ } & \colhead{($\mathrm{au}$)} & \colhead{$(\mathrm{au})$} & \colhead{$(\mathrm{km\,s^{-1}})$} & \colhead{ } & \colhead{ } & $(^{\circ})$ & \colhead{$(\mathrm{au})$} & \colhead{$(\mathrm{km\,s^{-1}})$} &  \colhead{$(\mathrm{km\,s^{-1}})$} & \colhead{$(\mathrm{km\,s^{-1}})$} \\
    \hline
  Low-velocity     & 200 &   0.996 & 0.618 & 4.670 & 0.307 & 1.151 & 135.710 & 4.478 & 1.363 & 8.854 \\
 (Red ellipses)    & 170 &  -0.524 & 0.554 & 4.621 & 0.296 & 1.168 & 131.305 & 4.415 & 1.400 & 8.091 \\
                   & 140 &  -4.328 & 0.560 & 4.543 & 0.292 & 1.240 & 130.207 & 4.317 & 1.450 & 8.023 \\
                   & 110 &  -7.308 & 0.549 & 3.962 & 0.279 & 1.593 & 126.806 & 3.696 & 1.451 & 7.865 \\
                   &  80 &  -1.056 & 0.548 & 3.361 & 0.248 & 2.224 & 118.629 & 2.985 & 1.557 & 7.860 \\
                   &  50 &   3.164 & 0.547 & 3.154 & 0.218 & 2.509 & 109.260 & 2.675 & 1.681 & 7.838 \\
                   & -50 & -11.280 & 0.554 & 3.147 & 0.228 & 2.538 & 112.881 & 2.693 & 1.636 & 7.940 \\
                   & -80 & -15.308 & 0.581 & 3.458 & 0.248 & 2.149 & 118.396 & 3.074 & 1.595 & 8.340 \\
                   &-110 & -25.300 & 0.574 & 3.902 & 0.275 & 1.672 & 125.868 & 3.615 & 1.489 & 8.227 \\
                   &-140 & -29.352 & 0.590 & 4.441 & 0.291 & 1.277 & 129.860 & 4.215 & 1.428 & 8.464 \\
                   &-170 & -19.540 & 0.603 & 4.727 & 0.301 & 1.144 & 133.220 & 4.522 & 1.413 & 8.640 \\
                   &-200 & -17.204 & 0.614 & 4.812 & 0.315 & 1.131 & 139.315 & 4.619 & 1.385 & 8.806 \\
    \hline 
  High-velocity    & 110 &  -4.684 & 0.788 & 7.056 & 0.193 & 0.085 & 81.641 & 7.063 & 0.383 & 11.293 \\
  (Brown ellipses) &  80 &  -2.996 & 0.921 & 6.173 & 0.158 & 0.158 & 67.226 & 6.153 & 0.664 & 13.206 \\
                   &  50 &  -4.444 & 0.764 & 5.544 & 0.135 & 0.177 & 57.561 & 5.514 & 0.698 & 10.952 \\
                   & -50 & -19.272 & 0.803 & 5.828 & 0.143 & 0.173 & 61.124 & 5.799 & 0.710 & 11.516 \\
                   & -80 &  -9.948 & 0.945 & 6.022 & 0.169 & 0.154 & 71.682 & 6.010 & 0.575 & 13.549 \\
                   &-110 &  -9.588 & 0.820 & 6.474 & 0.174 & 0.112 & 73.745 & 6.473 & 0.471 & 11.761 \\
    \hline 
    \end{tabular}
    \label{tab:parameters}
\end{table*}

The extended vertically distributed emission around $V_\mathrm{off}=0$ in Figure \ref{fig:pv_diskwinds} may be attributed to emission from the parent cloud or associated with  ambient material of the outskirts of the outflow (e.g., \citealt{Shang2023a}; \citealt{Ai2024}). The velocity gradient along the line of sight detected in the velocity map of Figure \ref{fig:moment1} is also observed in these PV diagrams, indicating substantial dominant ambient contribution to the blue side by preexisting cloud inhomogeneity. Note that the CO emission exhibits a possible shell structure consistent with radially expanding shells (\citealt{Arce2011}; \citealt{Zapata2014}), similar to the SO emission observed in Figure \ref{fig:pv_jetaxis}. This could be indicative of the elliptical structure expected in  the perpendicular PV diagrams of an outflow with a low inclination angle with respect to the plane of the sky (\citealt{Lee2000}).

To determine the kinematic and physical properties of the outflow, we performed cuts in each PV diagram at intervals of $0^{\prime\prime}.06$ (synthesized beam size). From these cuts, we extracted spatial profiles and identified the 2$\sigma$ limit, correlating these with positions in the PV diagram. Each position defines a point in the PV diagram. We then fitted ellipses to the points associated with the 2$\sigma$ limit following the CO and SO data in all PV diagrams for red ellipses (which exclude the high-velocity components) and for $z\leq|110|\,\mathrm{au}$ for the dashed brown ellipses (which concentrate on the higher-velocity portion of the data closer to the disk) of Figure \ref{fig:pv_diskwinds}. We do not fit ellipses to the higher-velocity emission at heights $z>|110|\,\mathrm{au}$ because the emission is dominated by the low-velocity components, making it complex to distinguish the boundary between the two. After the fitting, we obtained the parameters of the ellipse, position angle (PA), major ($a$) and minor ($b$) axes, and the coordinates of the center ($r_\mathrm{cent}$, $V_\mathrm{cent}$). Using the outflow model presented by \citet{Louvet2018}, we estimated the physical properties of the outflow, such as cylindrical radius ($R$), radial expansion velocity ($V_\mathrm{exp}$), rotation velocity ($V_\mathrm{rot}$), and subsequently, we estimate the specific angular momentum ($j_\mathrm{obs}$). The parameters obtained from our fits and the values of the $R$, $V_\mathrm{exp}$, $V_\mathrm{rot}$, and the axial velocity $V_z$ are shown in Table \ref{tab:parameters}. The expansion, rotation, and axial velocities are corrected for the inclination angle with respect to the line of sight $i\sim86^{\circ}$ (\citealt{Claussen1998}). 

The fitted ellipses are used to determine the physical properties depicted in Figure \ref{fig:dwparameters}. The red and brown sets of ellipses produce two parallel sets of data lines, displaying consistent differences and similarities. 
The top left panel displays the assumed cylindrical radius as a function of the height  for a conical shell. These radii increase with  height, indicating the expansion of the  shell. The top right panel of Figure \ref{fig:dwparameters} illustrates the expansion velocity, which also increases with height, as observed in PV diagrams depicted in Figure \ref{fig:pv_diskwinds}. The rotation velocity is depicted in the bottom left panel and decreases with height, consistent with findings from prior research by \citet{Lee2018}. The bottom right panel shows the specific angular momentum $j_\mathrm{obs}=R\times V_\mathrm{rot}$. The specific angular momentum from the red ellipses maintain constant at around 190 $\mathrm{au}\,\mathrm{km\,s^{-1}}$, while for the brown ellipses is around 50 $\mathrm{au}\,\mathrm{km\,s^{-1}}$. Error bars for these properties are estimated by propagating the statistical uncertainties derived from the ellipse fitting incorporated into the outflow model of \citet{Louvet2018}. Stronger than most of these $z$-dependence, are the consistent differences between the red and brown sets of ellipses. These show smaller $R$, $V_\mathrm{rot}$, and $j_\mathrm{obs}$ for the brown set. The smaller $V_\mathrm{exp}$ for the red set of ellipses is consistent with their choice of avoiding high-velocity data in the PV\@. 

Both sets of ellipses are subject to possible overestimates of rotation velocity $V_\mathrm{rot}$, and therefore the specific angular momentum $j_\mathrm{obs}$, because the ellipses method is only valid for a narrow emitting ring, given that, for radially extended disk winds, we observe the summed contribution of the a broad range of the rings. The poloidal and the vertical velocities present high uncertainties associated with a very high inclination angle of the jet axis to the line of sight.

Several PV diagrams shown in Figures \ref{fig:pv_xwinds} and \ref{fig:pv_diskwinds} may be contaminated by the emission from CO and SiO knots; however, this emission does not affect the estimation of the physical parameters of the outflow because the knots trace the emissions of higher intensity ($>10\sigma$) and higher velocity ($V_\mathrm{off}>|5|\,\mathrm{km\,s^{-1}}$) around the jet axis. Our estimations were based on the emission at $2\sigma$.

\begin{figure*}[t!] 
\centering
\includegraphics[width=\linewidth]{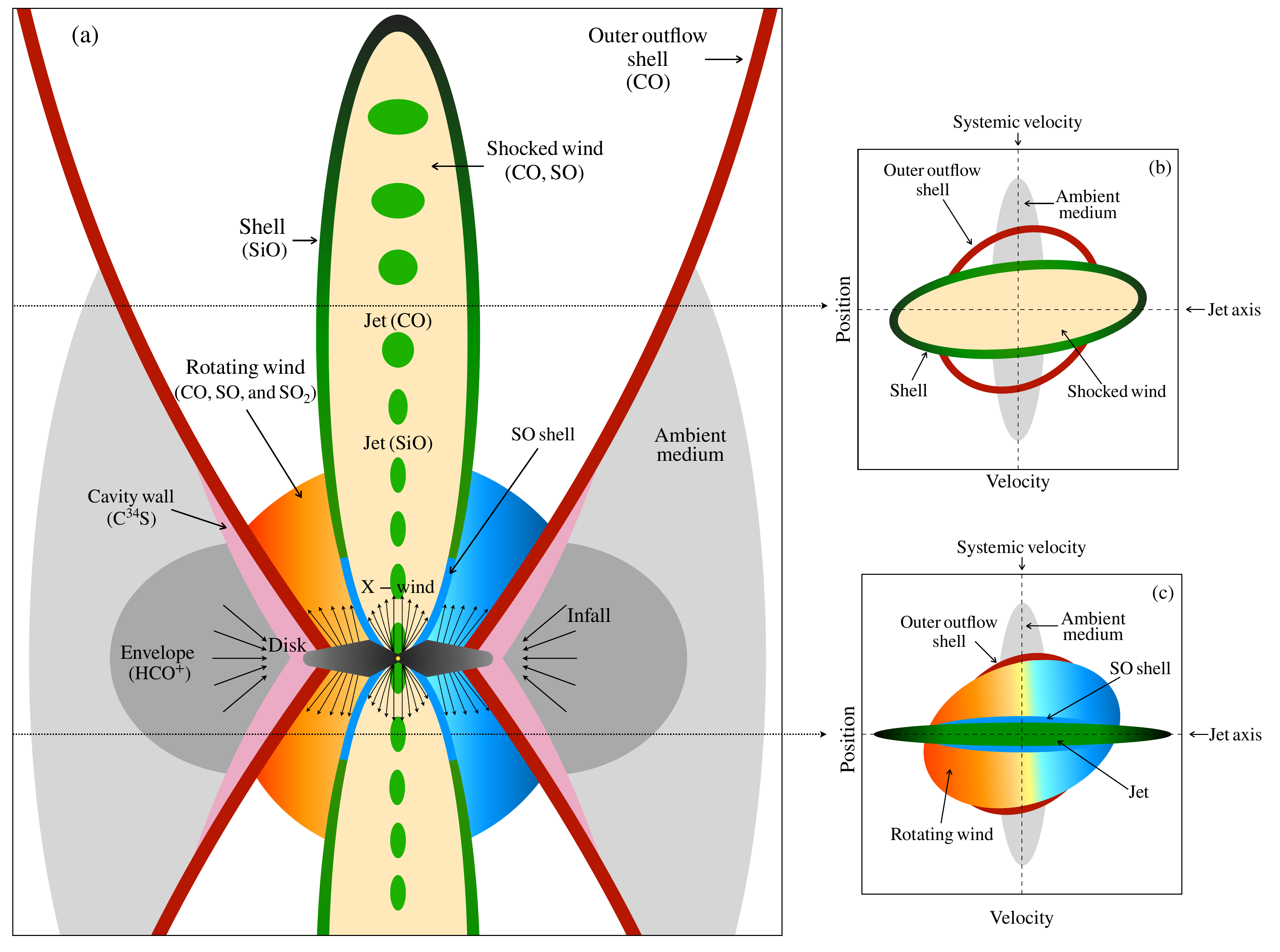}
\caption{Schematic diagram illustrating the components of the protostellar system of HH 212. The right panels depict PV diagrams perpendicular to the jet axis, showcasing various components of the system, including the ambient medium, envelope, cavity wall, an outer outflow shell, rotating wind, SiO shell, X-wind, and jet. The illustration is not to scale.}
\label{fig:cartoon}
\end{figure*}

\section{Discussion} \label{sec:discussion}

\subsection{Onion-like Spatial and Kinematic Structures}
\label{subsec:origin_onion}

Figure \ref{fig:cartoon}a summarizes the different components of the HH 212 protostellar system. Given that the outflow has a very low inclination angle with respect to the plane of the sky, the high angular resolution observations reveal a well-defined onion-like spatial structure. From outermost to axial regions, one can observe the infalling envelope (\citealt{Lee2014,Lee2017ApJ}), the cavity wall (\citealt{Codella2014}; \citealt{Tabone2017}), the outer outflow shell, the slow wind that is rotating, the SiO shell, the  axially extended shocked wind region, and the jet.

The CO outer shell depicted in Figure \ref{fig:cartoon}a is associated with the molecular outflow observed through CO molecular line emission (\citealt{Lee2015}). This shell, traced by CO, could result from  swept-up material by the outflowing gas from the star or the disk (e.g., \citealt{Snell1980}; \citealt{Shu1991}; \citealt{JALV2019}), or from the interaction between the wind launched very close to the central star or the jet with the ambient material, (e.g., \citealt{Li1996}; \citealt{Canto2006}; \citealt{Shang2023a}). Further discussion on the possible origins of these layers of gas will be provided later.

From the outer to inner regions, the system is characterized by the three dominant components: (1) the rotating wind region, traced by the emission of the CO and SO molecular lines, confined by an outer shell, and (2) the SiO shell, and (3) the CO shocked wind and the wide-angle component of the X-wind inside of the SiO shell.

The SiO shell on the parallel PV diagram, located between the wide-angle MHD wind and shocked wind, was identified by \citet{Lee2022}. The SiO shell on the PV diagram marks the separation between the large-scale elongated shocked region where CO and SO coexist, and the base SiO jet emission displaying large velocity widths arising in the wide-angle X-wind \citep{Lee2022}. It was proposed that the SiO shell could arise by the cumulative effect of several large jet bow shocks impacting an outer disk wind (\citealt{Lee2021}), or by a radial component of the wide-angle X-wind expanding into the extended disk wind. 
The presence of the extended SiO nebulous emission-filled structure is, in fact, consistent with the existence of the large shocked wind region surrounded and confined by the poloidal magnetic field in the compressed ambient region as shown in \citet{Shang2020,Shang2023b}. The large-scale axially extended SiO-filled shocked wind region is downstream of the reverse shock cavity, interfacing the newly launched wide-angle X-wind and the lateral disk wind region dominated by CO and SO\@.

Another explanation of the origin of the CO shocked wind inside of the SiO shell is material ejected sideways from the CO and SiO knots. \citet{Tabone2018} explored the interaction between a pulsating jet and a surrounding disk wind. In their simulations, the perturbations of the disk wind, caused by bow shocks are confined inside a cone, while the gas outside of this cone remains unperturbed disk wind material.

Figures \ref{fig:molecules_resolution}b and \ref{fig:moment1}b show that the CO molecular line traces the same spatial region as the SO\@. In Figure \ref{fig:pv_jetaxis}c, CO traces the material within the expanding shell on the PV diagram. Therefore, it can be inferred that CO in a region close to the central source, $z\leq |200|\,\mathrm{au}$, could also be associated with the emission of the disk winds. The possible shocked CO emission is primarily characterized by the lower-$J$ CO transitions (3-2) in Band 7. The  extended and elongated axial region, confined by the SiO shell, contain CO bullets and emission SO. This extended shocked wind region likely accumulates the shocked wind material accumulated from the previously shocked wide-angle portion of the X-wind (\citealt{Lee2022}) interacting with the surrounding ambient environment, then advected downstream. Outside the SiO shell, and exterior to the base SO shell (lower cavity), CO, SO and SO$_2$ emission coexist in an horizontally extended region bounded from outside by the outer CO shell. Connecting the two upper and horizontal shocked regions are CO bridges crossing the SiO and SO shell-like structures.

The inner region is dominated by the emission of the episodic jet. \citet{Lee2017Nat} identified a chain of SiO rotating knots associated with the jet. These knots are situated close to the accretion disk, $z\leq |80|\,\mathrm{au}$. In contrast, we detect a chain of knots traced by CO emission, as shown in Figures \ref{fig:zoom_knots} and \ref{fig:knots_position}. However, the CO knots are located at a greater distance compared to the SiO knots, $|100|\,\mathrm{au}\leq z\leq |700|\,\mathrm{au}$. The width of the CO knots, similar to that of the SiO knots, increases with distance from the central source. This may be due to a decrease in pressure along the jet axis, causing the knots to rapidly expand (\citealt{Smith1997}; \citealt{Suttner1997}),  or traced by larger blobs of compressed or shocked material advected or entrained from the outer interacting region (\citealt{Wang2019}).

The spatial onion-like structure is also observed in the PV diagrams. Along the jet axis (Figure \ref{fig:pv_jetaxis}), the PV diagram reveals that the SiO emission, both spatially and in velocity range, encompasses the emission of CO and SO.

Figure \ref{fig:cartoon}b illustrates the structure observed in PV diagrams of Figure \ref{fig:pv_xwinds}, wherein CO traces a slow outer shell, while the SiO delineates a faster internal shell. Inside the SiO shell, the emission is dominated by CO emission, which traces a shocked wind. Additionally, close to the central star, the PV diagrams (Figures \ref{fig:pv_xwinds} and \ref{fig:cartoon}c) show the emission of rotating winds traced by SO (\citealt{Lee2021}) and CO, where SO is a shock tracer. In this region, a velocity gradient along the line of sight is observed, likely associated with rotation of the gas. Furthermore, the SiO emission predominantly comes from the jet, exhibiting a compact structure with higher velocities respect to the emission of the SO and CO. 

The origin of the onion-like structure may be linked to the temperatures and critical densities of the molecular lines involved. For instance, the SiO jet has a temperature of 150 K and a critical density of $1.7\times10^6\,\mathrm{cm}^{-3}$ (\citealt{Dutta2024}). Its emission is associated with dense gas, such as the knots in the jet and the shell formed by collisions between the wide-angle component of the X-wind with the disk wind.

The CO line observed in the shell and in the rotating wind might have a temperature of 50 K and a critical density of $1.2\times10^{4}\,\mathrm{cm}^{-3}$ (\citealt{Dutta2024}), while CO associated with the shocked wind and with the jet has a temperature of 150 K and a critical density of $1.1\times10^4\,\mathrm{cm}^{-3}$. The distinction between these CO regions lies in the fact that the emission from the shell could be associated with entrained material by an outflowing gas, whereas the emission from the jet and the shocked wind could be heated by internal shocks between the launched gas and the jet. This explains why the shocked wind exhibits emission of the CO (3-2) (this work) and CO (6-5) (\citealt{Codella2023}). 

Finally, SO traces shocked gas by interactions, having a temperature ranging from 50 to 100 K and a critical density of $2\times10^{6}\,\mathrm{cm}^{-3}$ (\citealt{Podio2015}). As such, SO may trace the interaction between the inner wide-angle portion of the X-wind, with a disk wind. We discuss in the following sections.

\subsection{Origin of the CO Emission and the Molecular Outflow}
\label{subsec:origin_outflow}

As mentioned above, CO molecular line emission is observed in various regions of the protostellar system HH 212, including the outer outflow shell, the rotating wind, the shocked wind, and the jet/X-wind. CO emission is typically associated with the molecular outflow, so we will discuss the origin of both together. Notice that, the CO emission associated with the CO shocked wind and the jet was already discussed in \S\ref{subsec:origin_onion}.

The origin of the various shells can be explained in several ways. \citet{Lee2006} found that the morphological relationship between the H$_2$ jet and the CO outflow and the kinematic behavior of the CO outflow along the jet axis, supports the jet-driven bow shock model. Additionally, \citet{Lee2015} showed that the CO shell is associated with knots and bow shocks driven by them, suggesting that sideways material ejections from knots within the outflow cavity form bow shocks and internal shells.

Furthermore, \citet{Rabenanahary2022} described that this multiple shell structure observed in this molecular outflow can be explained as a result of the interaction between a narrow jet with a stratified core. The main difference between this interaction and that of the jet with uniform core is that the outflows open much wider. Additionally, after several hundred years, the initial shell splits into a slow parabolic outer shell and an inner, faster shell tracing the jet-ambient interface; eventually these shells stop expanding at the base, and reach a maximum width and opening angle in excellent agreement with observations.

Another possible explanation for the origin of the outflow shell is that it results from entrained material from the envelope by the outflowing gas from the protostar-disk system. \citet{JALV2019} modeled a molecular outflow as a dynamical shell produced by the collision between an anisotropic stellar wind and a rotating cloud envelope in gravitational collapse (\citealt{Ulrich1976}). In their model, the shell is fed by both the envelope and stellar wind, such that the mass and rotation velocity coming from the envelope, while the radial expansion results from the pushing effect of the outflowing gas.

However, the observed rotation velocity ($\sim 0.4$--$2\,\mathrm{km\,s^{-1}}$) is between two and ten times larger than the expected rotation velocity ($\sim0.1-0.2\,\mathrm{km\,s^{-1}}$) of the rotating envelope according to \citet{Ulrich1976},  as previously reported in Orion Source I \citep{JALV2020}. Therefore, it is plausible to consider that this faster rotation or asymmetry may arise from either an MHD disk wind (see \citealt{Tabone2017,Tabone2020,Lee2021}) a possibility further discussed in \S\ref{subsubsec:diskwind}, or an envelope rotating faster than the Ulrich rate \citep{Mendoza2009} that mixes with shocked X-wind material through magnetized interaction in the region between the outer shell and the reverse shock, a possibility further discussed in \S\ref{subsubsec:unifed}. In all of these cases, the CO gas originates from the envelope.

In the region close to the central source, $z\leq |200|\,\mathrm{au}$, between the  outer CO shell and the SiO shell, we observe that the CO and SO lines trace the rotating wind. This wind was discovered through SO and SO$_2$ emission by \citet{Tabone2017}. Subsequently, \citet{Tabone2020} modeled the emission of this wind and assessed it to be consistent with MHD disk wind solutions, which are used for extracting the excess of the angular momentum from the accretion disk. On the other hand, \citet{Lee2018} found SO expanding shell. 
Through high angular resolution observations, \citet{Lee2021} confirmed the existence of this shell, which represents the inner limit where the SO shocked wind may exist. This shell has a inner boundary of $4-8\,\mathrm{au}$, while the disk wind is expected to launch from a range of $4-40\,\mathrm{au}$.

Two parameters to describe the MHD disk winds are the launching radius $R_\mathrm{launch}$ and the magnetic lever arm parameter $\lambda$. The $\lambda$ parameter describes the magnetic torque exerted by the disk wind (e.g., \citealt{Pudritz2007}; \citealt{Alexander2014}; \citealt{Pascucci2022}).

\subsubsection{Free Ballistic Disk Wind Scenarios}
\label{subsubsec:diskwind}

\begin{figure}[t!]
\centering
\includegraphics[width=\linewidth]{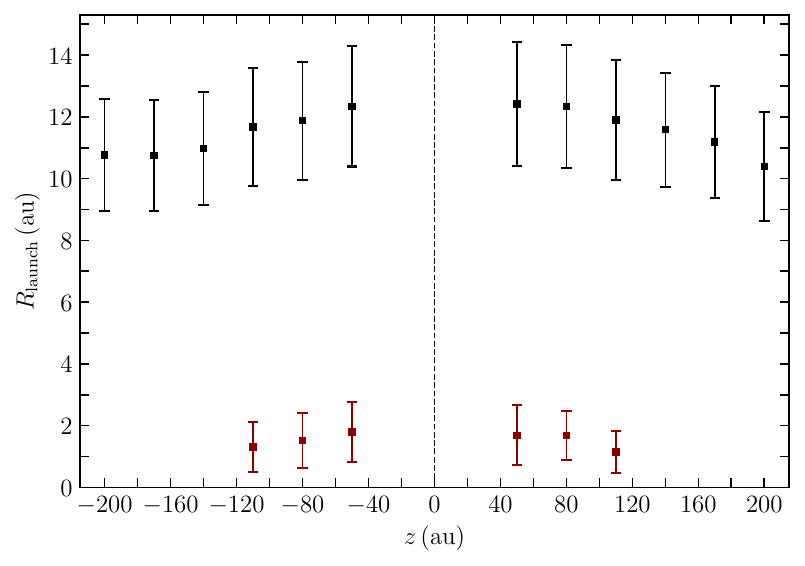}
\caption{The launching radii of the HH 212 CO wide-angle MHD disk winds as a function of the height respect to the disk mid plane $z=0$. The radii are estimated using Anderson's relation (\citealt{Anderson2003}), and the error bars represent the uncertainties in our estimations.}
\label{fig:shell_launch}
\end{figure}

\citet{Anderson2003} found a general relation between the velocity components of the wind at large distances and the rotation rate of the launching surface under following assumptions: (1) the wind is free, meaning that the wind is not interacting with other components of the protostellar system; (2) dynamically cold, such that the entropy effect is negligible; (3) the system is axisymmetric; (4) the system is in steady state and (5) the observation point is sufficiently distant from the launching point.

In the case of HH 212 protostellar system, the presence of the CO and SiO knots, the SiO shell, and the CO shell indicate that the disk wind is in constant interaction with the different components of the system and with itself. Thus, the wind is not free. Nevertheless, \citet{Tabone2018} described that the disk wind outside of the cone is unperturbed; therefore, it is possible to construct a free-wind model that fits the known observations. We connect the CO molecular line emission with the tracing of a MHD wind.

The following equation \citep{Anderson2003} allows to find the launching point of a free magnetocentrifugal MHD wind under the conditions of application mentioned above:
\begin{equation}
    \label{eq:anderson}
    V_\mathrm{outflow}^2R_\mathrm{launch}^{3/2}+\frac{3}{2}\eta^2 R_\mathrm{launch}-2 R V_\mathrm{rot}\eta \approx 0,
\end{equation}
where $V_\mathrm{outflow}=\sqrt{V_\mathrm{p}^2+V_\mathrm{rot}^2}$, $V_\mathrm{p}=\sqrt{V_z^2+V_\mathrm{exp}^2}$ is the poloidal velocity determined by the radial expansion velocity $V_\mathrm{exp}$ and the axial velocity along the jet axis $V_z$, $R$ is the cylindrical radius, and $\eta=(GM_*)^{1/2}$, where the mass of the protostar is $M_*=0.25\pm0.05$ $\mathrm{M_\odot}$ (\citealt{Codella2014,Lee2014}). The result of solving this equation is presented in Figure \ref{fig:shell_launch}, where we observe that the launching radii for the red ellipses are $\sim9 - 15\,\mathrm{au}$ and for the brown ellipses they are $\sim1-2\,\mathrm{au}$. The red-ellipse values are consistent with those reported by \citet{Tabone2020} and \citet{Lee2021} detail comparison with synthetic PVs for MHD disk wind models, where they estimated launching radii between $4$--$40\,\mathrm{au}$, while the brown ellipse values are connected to larger expansion velocities.

\begin{figure}[t!]
\centering
\includegraphics[width=\linewidth]{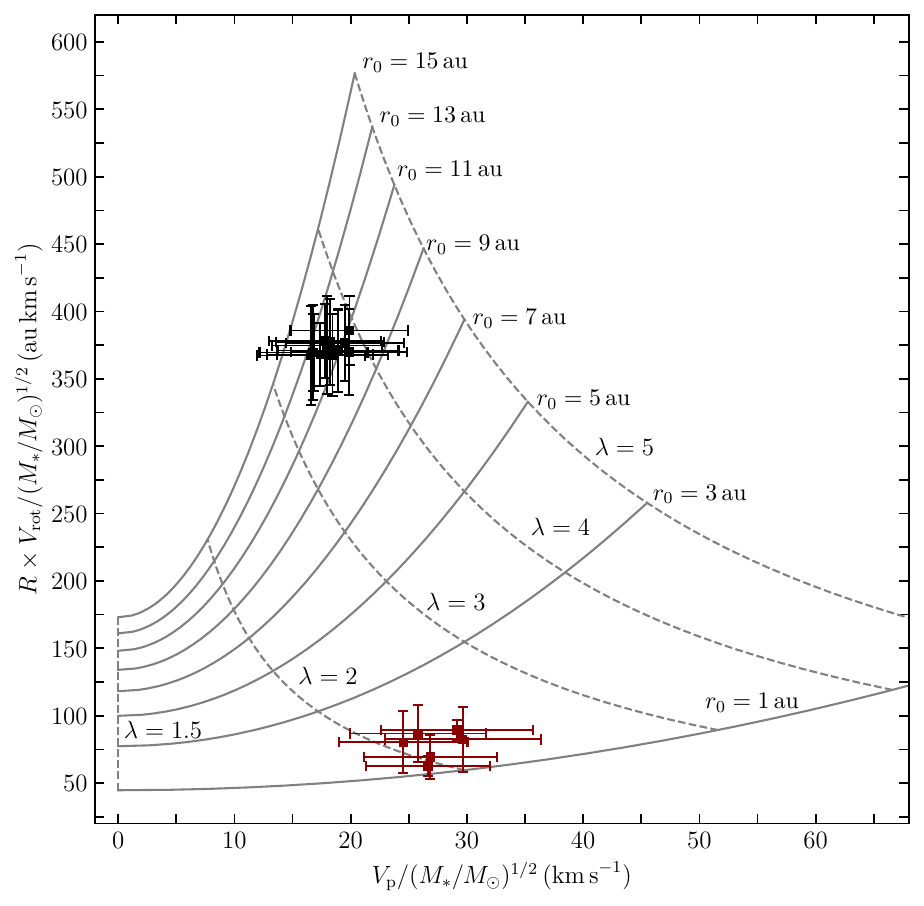}
\caption{The specific angular momentum as a function of the poloidal velocity, both normalized to $\sqrt{M_*}$. The black and brown symbols represent the estimated values for the HH 212 CO wide-angle MHD disk winds, while the curves depict the expected relation from self-similar cold magneto-centrifugal disk winds with launching radii $r_0$ from 1 to 15 au (solid lines) and magnetic lever arm parameter $\lambda$ from 1.5 to 5 (dashed lines). Figure adapted from \citet{Ferreira2006}.}
\label{fig:shell_lambda}
\end{figure}

As mentioned above, another parameter in the MHD disk winds is the magnetic lever arm parameter $\lambda\sim(r_\mathrm{A}/r_0)^2$, where $r_\mathrm{A}$ is the cylindrical radius at the Alfv\'en surface and $r_0$ is the launching point of the streamline that follows the wind. In previous works (\citealt{Tabone2017,Tabone2020,Lee2021}), magnetic lever arm parameter of the SO disk wind was found to be $3.5$--$5.5$. 
We estimate the parameter $\lambda$ value of our CO observations in the asymptotic regime ($R/r_0\to \infty$) where the specific angular momentum extracted by the magnetic torque has been converted in matter rotation (\citealt{Blandford1982}). The asymptotic values of the poloidal velocity and the specific angular momentum at different heights $z$ are (\citealt{Blandford1982}):
\begin{equation}
    \label{eq:vpol}
    V_\mathrm{p}=\sqrt{2\lambda-3}\sqrt{G\mathrm{M_*}/r_0},
\end{equation}
\begin{equation}
    \label{eq:vrot}
    R\times v_\mathrm{rot}=\lambda \sqrt{G \mathrm{M_*} r_0}.
\end{equation}
Figure \ref{fig:shell_lambda} shows the relation between the specific angular momentum $j=R\times v_\mathrm{rot}$ and the poloidal velocity $V_\mathrm{p}$ for the various solutions of the launching point $r_0$ and magnetic lever arm parameter $\lambda$. Based on our CO observations, we   find that for the red ellipses the launching points are $r_0\approx9-15$ and the magnetic lever arm parameter is $\lambda\sim3-4.5$. These values are consistent with those previously reported for SO observations. However, for the brown ellipses, that do avoid the strong envelope inhomogeneities of the PV diagrams, the picture changes, giving a launching point $r_0\sim1\,\mathrm{au}$, substantially smaller, and $\lambda\sim2-2.5$.

Regarding the red ellipses, given that the launching point and the magnetic lever arm parameter estimated for the CO molecular line are within the range of previously reported values, we conclude that the CO in the region between the CO shell and SiO shell can be parameterized by means of a MHD disk wind launched from the region $\sim15$--$40\,\mathrm{au}$. The lower limit of $\sim9$--$15\,\mathrm{au}$ likely marks the boundary where the MHD disk wind interacts with the X-wind, as traced by the emission of the SO (\citealt{Lee2021}), while 40 au is the disk radius (\citealt{Lee2017ApJ}). The discrepancy in the lower limit of 4 au for the wind launching point reported by \citet{Lee2021} arises because our velocities and radii measurements were made at the emission limit (2$\sigma$), whereas \citet{Lee2021} determined the physical properties at the peak of the SO emission. We choose to measure the physical quantities at 2$\sigma$ to avoid contamination from the CO shocked wind inside the SiO shell (Figures \ref{fig:molecules_resolution}, \ref{fig:moment0}, and \ref{fig:pv_jetaxis}) and SO shell (see Figure \ref{fig:pv_jetaxis}). The multiple shell structure observed in the molecular outflow of HH 212 could be produced by a pulsing jet or wind propagating into a uniform ambient medium (see \citealt{Lee2001}).

The outflow associated with HH 212 exhibits characteristics of both jet and wide-angle wind features. Therefore, we suggest that both effects coexist and play decisive roles in the formation and evolution of molecular outflows. At this juncture, the unified jet-wind models such as in \citet{Shang2006} are a direction promising to explain the dual characteristics.

\subsubsection{The Unified Model}
\label{subsubsec:unifed}

In recent numerical simulations, \citet{Shang2020,Shang2023a} propose a unified model where molecular outflows arise from the magnetic interaction between the ambient medium and the X-wind launched from the disk. In this model, the outflow is formed by a magnetized primary wide-angle X-wind that bears a collimated high-density jet along the outflow axis. The interaction between the wide-angle wind and the ambient envelope creates a bubble structure with  nested layers of structures, with an extended shocked region of mixing of wind and ambient materials between the forward and reverse shocks, and the free X-wind, exists inside the  reverse shock cavity.  Due to the magnetic interplay between the X-wind and the envelope, multi cavities looking like multiple large ejections can appear as a result, which resembles the multiple nested components observed in this study.

The structure proposed in the unified model is observed in the protostellar system HH 212. From the outermost to innermost structures relative to jet axis, the cavity wall traced by C$^{34}$S (\citealt{Codella2014}; \citealt{Tabone2017}) represents the shocked ambient medium, corresponding to the material between the forward shock and the contact discontinuity. The CO shell and the SiO shell denote the high-density shocked wind outlining the boundaries of the multicavities. Meanwhile, the wide-angle MHD disk wind and shocked wind of the CO represent the low-density shocked wind. Finally, the CO emission observed between the SiO knots (Figure \ref{fig:molecules_resolution}c) is associated with the large pseudopulses, caused by oscillations in magnetic forces \citep{Shang2020}.

It has already been proven that the rotating jet derives from an X-wind model (\citealt{Lee2017Nat}). The most straightforward interpretation of the facts of this outflow is more naturally unified with this known fact, in the form of the unified model. The unified model explains the features of the outflow as the result of the magnetic interaction between the wind launched from the protostellar-disk system and the ambient medium. The observed rotation in the rotating region of HH 212 can be attributed within the unified model to the inhomogeneity of outflow lobe in the $\phi$-direction, or to the incorporation of angular momentum of the infalling and collapsing ambient cloud matter entrained through the Kelvin--Helmholtz instability. Future very high-angular resolution observations of molecular outflows with different inclination angles with respect to the line of sight and new analyses that consider rotation in the envelope will be crucial for elucidating the detailed structures produced by the magnetized bubble driven by the unified model, with the hope of discriminating from the free disk wind scenarios.

\section{Conclusions} \label{sec:conclusions}

We provide a comprehensive analysis of the ALMA observations of the CO line emission in the protostellar system HH 212. We also compare the results with the previous results of the SiO and SO line emission to have a more complete picture of the outflow.

The CO molecular line emission delineates four distinct regions: an outer outflow shell, the shocked rotating wind between the CO and SiO shells, the shocked wind within the SiO shell, and the jet/X-wind.

The CO outer outflow shell is the one previously reported by \citet{Lee2015}. This shell is the result of the interaction between the rotating wind with the ambient medium and shows a dominance and influence from the surrounding envelope.

We find that the low velocity components of CO rotating wind can be tracing an expanding shell launched from a disk radius of $\sim 9-15\,\mathrm{au}$. This launching point is larger than that of the previously detected SO shell, suggesting that the CO shell likely marks the boundary where the wide-angle component of the X-wind interacts with the MHD disk wind. This CO wind has a magnetic lever arm parameter of $\lambda \sim 3-4.5$. 

Under assumption that the high velocity components of the rotating wind is tracing a MHD disk wind, this emission could be parameterized with launching radii $\sim 1-2\,\mathrm{au}$ with a magnetic lever arm parameter of $\lambda \sim 2-2.5$. However, this emission is likely associated with a shocked wind, as a result of the interaction between the jet or wide-angle X-wind with the surrounding, the envelope, disk atmosphere, or a disk wind. Also, this emission could result from material ejected sideways by the internal bow shocks driven by the CO or SiO knots.

Additionally, the CO jet exhibits discernible knots distributed at roughly equal intervals from the central source, $|100|\,\mathrm{au}<z<|700|\,\mathrm{au}$. Furthermore, the width of the knots increases with the distance away from the central source.

Finally, the shocked CO wind coexists with SO inside the large elongated shell on top of the stream of knots far downstream, converging from the outer shell midway of the SiO. This CO wind could be associated with the wide-angle component of the X-wind shocked in the compressed wind region or entrained from the compressed ambient medium.

\begin{acknowledgments}
We thank an anonymous referee for very useful suggestions that improved the presentation of this paper.
We are grateful to Susana Lizano, Jorge Cant\'o, Manuel Fern\'andez-L\'opez for the discussion and helpful comments about the outflow models. We are grateful with Benoit Tabone for his contribution in the ALMA proposal \#2016.1.01475.S.
J.A.L.V. and C.F.L. acknowledge grants from the National Science and Technology Council of Taiwan (NSTC 112–2112–M–001–039–MY3) and the Academia Sinica (Investigator Award AS--IA--108--M01). 
H.S. acknowledges support from the National Science and Technology Council in Taiwan (NSTC 113-2112-M-001-008). 
LP and CC acknowledge financial support under the National Recovery and Resilience Plan (NRRP), Mission 4, Component 2, Investment 1.1, Call for tender No. 104 published on 2.2.2022 by the Italian Ministry of University and Research (MUR), funded by the European Union–Next Generation EU-Project Title 2022JC2Y93 Chemical Origins: linking the fossil composition of the Solar System with the chemistry of protoplanetary disks – CUP J53D23001600006 – Grant Assignment Decree No. 962 adopted on 30.06.2023 by the Italian Ministry of Ministry of University and Research (MUR). LP and CC also acknowledge the PRIN-MUR 2020 BEYOND-2p (Astrochemistry beyond the second period elements, Prot. 2020AFB3FX), the project ASI-Astrobiologia 2023 MIGLIORA (Modeling Chemical Complexity, F83C23000800005), the INAF-GO 2023 fundings PROTO-SKA (Exploiting ALMA data to study planet forming disks: preparing the advent of SKA, C13C23000770005), and the INAF Mini-Grant 2022 ``Chemical Origins'' (PI: L. Podio).
A. M. acknowledges the grant from the National Science and Technology Council (NSTC) of Taiwan 112-2112-M-001-031- and 113-2112-M-001-009-.
This paper makes use of the following ALMA data: ADS/JAO.ALMA \#2015.1.00024.S, ADS/JAO.ALMA \#2015.1.00041.S, and ADS/JAO.ALMA \#2016.1.01475.S. ALMA is a partnership of ESO (representing its member states), NSF (USA) and NINS (Japan), together with NRC (Canada), NSTC and ASIAA (Taiwan), and KASI (Republic of Korea), in cooperation with the Republic of Chile. The Joint ALMA Observatory is operated by ESO, AUI/NRAO and NAOJ.
\end{acknowledgments}

\bibliography{sample631}{}
\bibliographystyle{aasjournal}




\end{document}